# Estimating Interaction Strengths with Correlations in Annual Relative Weight: Interspecific Competition and Predation in Fishes, Pueblo Reservoir, Colorado

Joshua M. Courtney,[1] Amy C. Courtney,[1] and Michael W. Courtney[2]

[1]BTG Research, P.O. Box 62541, Colorado Springs, CO, 80962
[2]United States Air Force Academy, 2354 Fairchild Drive, USAF Academy, CO, 80840
Michael_Courtney@alum.mit.edu

**Background:** Quantifying the interaction strength between species is of general interest in ecology and food web studies for understanding system stability and for predicting system responses to perturbations. Theory has run far ahead of experiment in solving equations describing ecological systems (the Lotka-Volterra equations, for example). Many modeling approaches depend upon quantitative knowledge regarding interaction strengths among species, yet there are very few methods available for estimating the strength of interactions between and among species, especially in the field. Interspecific competition and predation in fishes is traditionally studied with a variety of methods, most of them requiring extensive sampling to determine stock densities or extensive stomach content or isotope analysis. Average relative weight of fish species in an ecosystem can usually be accurately determined with much smaller sample sizes than required for accurate determination of stock densities.
**Materials and Methods:** Survey data was obtained from Colorado Division of Parks and Wildlife for thirteen species of fish in Pueblo Reservoir from 1990 to 2011. Relative weights were computed for each species in years with adequate sample sizes. Correlations of annual mean relative weights were then computed for each pair of species and for different length classes of a given species in some cases.
**Results:** When two species have a strong predator-prey relationship, correlation of annual mean relative weights tends to be strongly negative. One possible interpretation is that in years with large numbers of prey species, the prey species tends to be thin, because intraspecific competition depletes the food resources of the prey species. However, an abundant quantity of prey provides ample food which tends to increase the mean relative weight of the predator species. In contrast, in years with relatively few numbers of the prey species, the prey species tends to be plump, because their relative weight is not limited by available forage and intraspecific competition. When two species compete for the same food resources, their relative weights tend to be positively correlated, because they both tend to be plump in years when their shared forage resources are abundant, and they both tend to be thin in years when their shared food resources are scarce. In cases where the prey are dominated by the age zero cohort and the relative weights were determined in adults, the correlation relationships tend to be reversed. This gives rise to the fecundity hypothesis: high relative weights among the adults of a species give rise to strong age zero cohorts which lead to positive correlations between mean annual relative weights of predator and prey species. The strength of correlations varies with length class in a predictable way.
**Conclusion:** It might be an overstatement to suggest correlations of annual relative weights could be used isolation to test hypotheses and establish interaction strengths regarding competition and predation among fishes, especially since there are some cases of strong positive and strong negative correlations where corresponding predation and competition relationships are not independently established. However, the trends are sufficiently suggestive to use a strong positive correlation as supporting evidence of competition and possibly indicative of the magnitude of competition if other evidence such as analysis of stomach contents is also present. Likewise, a strong negative correlation might be interpreted as relevant when forming a hypothesis of predation and as supporting evidence of the magnitude of predation if other evidence is also present. In systems where regular weight and length survey data are available, this method can suggest the likely strength of competitive and predatory relationships and may be more cost effective than stomach content or isotope analysis. In systems with sufficient length-weight data, this method may be useful for quantifying the relative interaction strengths of predatory and competitive relationships between species.

**Key Words**: food web modeling, interaction strengths, population dynamics, habitat complexity, relative weight, interspecies competition, Lotka-Volterra equations, interaction matrix, predator-prey, walleye, common carp, white sucker, gizzard shad, fecundity hypothesis

**Introduction**
Predictive modeling of population dynamics in ecosystems requires quantitative information regarding the interaction strengths of species demonstrating significant ecological relationships. However, at present, there are very few field systems where these interaction strengths are sufficiently well known to actually study the predictive time evolution of systems of differential equations believed to describe the system (Wootton and Emmerson, 2005). There are no generally applicable methods for quantifying interspecific interaction strengths (Berlow et al., 2004). As a result, most "predictive" multispecies modeling focuses on the question of qualitative stability rather than making quantitative predictions regarding populaiton dynamics (May, 1973). Solving the Lotka-Volterra equations (or any analogous model) requires populating an interaction matrix which quantifies the interaction strength between every possible interacting pair of species in the system. Realistic predictive modeling is further complicated by the possibility of size (or life stage) dependent interaction strengths (Pimm and Rice, 1987). "The ecological community urgently needs to explore new ways to estimate biologically reasonable model coefficients from empirical data ..." (Berlow et al., 2004). The present study suggests the possibility of using correlations of mean relative weight in fish to assess the interaction strength of aquatic species, and by extension, the possibility of using analogous metrics of body condition in other species. This paper presents empirical results which may be usefully transformed into interaction strengths for predictive models; however, we leave suggestions on how to adapt our results and method to theoretical modeling for future study.

Relative weight is the ratio of a fish's actual weight with the standard weight of a fish of the same length, times 100 (Anderson and Neumann, 1996). Based on the premise that the standard weight, $W_s$, should represent fish in better than average condition, it is developed with the goal of representing the 75$^{th}$ percentile weight at a given length over a wide range of populations (Blackwell et al., 2000). Thus relative weight is an inherent measure of fish plumpness, and it is also commonly used to assess prey availability, evaluate management decisions, and compare with other populations (Blackwell et al., 2000).

Traditional approaches to studying predator-prey relationships in fishes are stomach content analysis and analyses of stock density indexes (Anderson and Neumann, 1996; Ney, 1990; Liao et al., 2001). The available prey to predator ratio describes the available prey crop (biomass per surface area of available prey) per unit of the cumulative predator crop (biomass per surface area of the predators). Kohler and Kelly (1991) point out that the most efficient way to assess the adequacy of the forage base is to sample the predators, because the condition of the predators is a reflection of forage supply over time.

The approach here extends that reasoning to multiple trophic levels because the prey species at one trophic level is often a predator at a lower level. Because a predator species at a relatively high trophic level depends on a prey species at a lower level, it will tend to be plump when its prey are relatively abundant compared with the demand for that prey. However, when the prey species are relatively abundant, then its own forage base is likely to be stressed due to intraspecific competition. Consequently, there is an expectation that the relative weight of predator and prey species tend to be negatively correlated.

This negative correlation is well known in small impoundments with relatively few fish species and relatively simple interaction dynamics (Anderson and Neumann, 1996). For example, the challenge in maintaining a stable equilibrium with good production of both bluegill and largemouth bass in farm ponds is well known. Management favoring largemouth bass production can tip the balance to large numbers of small bluegill, but then the bluegill tend to have a smaller proportional stock density (PSD 10-50) than largemouth bass (PSD 50-80). Conversely, management can favor production of desirable bluegill by tipping the balance to large numbers of smaller largemouth bass to keep the population of smaller bluegill in check to allow the bluegill that survive to larger size classes to be plump and desirable to panfish anglers. When managed for panfish, the PSD of bluegill might be in the range of 50-80; whereas, the PSD of largemouth bass would be closer to 20-40 (Willis et al., 1993).

In principle, one could study relative weight correlations from either multi year data or from data including multiple lentic ecosystems. The approach of using multiple systems is a bigger challenge because of the wider variety of species mixtures present in different bodies of water and the resulting greater differences in predator-prey interactions. Studying relative weight correlations over many years in a single body of water has the advantage of less variations in the species present, though some variation occurs in stocking rates, reproduction rates, and recruitment.

Pueblo Reservoir is a reservoir at 4900 feet above sea level on the Arkansas River of 5600 surface acres at full pool and a maximum depth of about 70 feet. It's capacity is approximately 350,000 acre feet. Detailed water quality, stream flow, and hydrodynamic data are available from several sources (Edlemann et al., 1991; Galloway et al., 2008; Lewis et al., 1994). Pueblo Reservoir is regularly stocked with black crappie (*Pomoxis nigromaculatus*), channel catfish (*Ictalurus punctatus*), rainbow trout (*Oncorhynchus mykiss*), largemouth bass (*Micropterus salmoides*), walleye (*Stizostedion vitreum vitreum*), and hybrid striped bass (*Morone saxatilis X chrysops*). Other species that present in the reservoir (through natural reproduction and/or residual from past stockings) in sufficient numbers to show up regularly in survey data are common carp (*Cyprinus carpio*), bluegill (*Lepomis macrochirus*), gizzard shad (*Dorosoma cepedianum*), smallmouth bass (*Micropterus dolomieui*), spotted bass (*Micropterus punctulatus*), white crappie (*Stizostedion vitreum vitreum*), white sucker (*Catostomus commersoni*), and yellow perch (*Perca flavescens*). This potpourri of cool and warm water fishes results from a mid-elevation reservoir that does not freeze over most years, and an angling public enthusiastic about cool water species.

This paper reports the results of computing year to year mean relative weights and correlations between different species of annual mean relative weights. In some cases, the correlations can be interpreted as relating to interspecies predation or competition. In other cases, the interpretation of the correlations remains unclear.

**Method**

Weight and (total) length survey data from 1990 to 2011 was provided by the Colorado Division of Parks and Wildlife. Fish were sampled with a combination of methods: seine, gill net, trap net, and electrofishing. For each data point including weight and length where the total length was above the appropriate minimum length, relative weights were computed from appropriate standard weight equations from Anderson and Neumann (1996) or Murphy et al. (1991) for fish not included in Anderson and Neumann. Mean relative weights were computed for each year, along with the standard error of the mean as the estimated uncertainty.

Table 1 shows the thirteen species and hybrids for which there was sufficient survey data for inclusion, along with the three letter code used by Colorado Division of Parks and Wildlife to identify each species and hybird and the a and b weight length parameters used for the standard weight. Figure 1 shows graphs of the standard weight equations for the fish considered here. Standard weight equations are plotted from the lower limit of their applicability up to the largest fish of each species in the data set for Pueblo Reservoir.

Once the mean annual relative weight was computed for each species and hybrid for each year with available data then interspecies correlation coefficients were computed for each pair of species and species hybrid combinations.

Table 1: Thirteen species and hybrids included in this study, along with three letter codes, and standard weight parameters a and b. The a and b parameters are for metric units (weight in grams, total length in mm), and the parameter a is the vertical intercept resulting from the linear least squares regression of log(W) vs. log(TL) so that the standard weight is $W_s = 10^a (TL)^b$.

| Species/Hybrid | Code | a | b |
| --- | --- | --- | --- |
| Bluegill | BGL | -5.374 | 3.316 |
| Channel Catfish | CCF | -5.800 | 3.294 |
| Black Crappie | BCR | -5.618 | 3.345 |
| Common Carp | CPP | -4.418 | 2.859 |
| Gizzard Shad | GSD | -5.376 | 3.170 |
| Rainbow Trout | RBT | -4.898 | 2.990 |
| Smallmouth Bass | SMB | -5.329 | 3.200 |
| Spotted Bass | SPB | -5.392 | 3.215 |
| Wiper | SXW | -5.201 | 3.139 |
| Walleye | WAL | -5.453 | 3.180 |
| White Crappie | WCR | -5.642 | 3.332 |
| White Sucker | WHS | -4.755 | 2.940 |
| Yellow Perch | YPE | -5.386 | 3.230 |

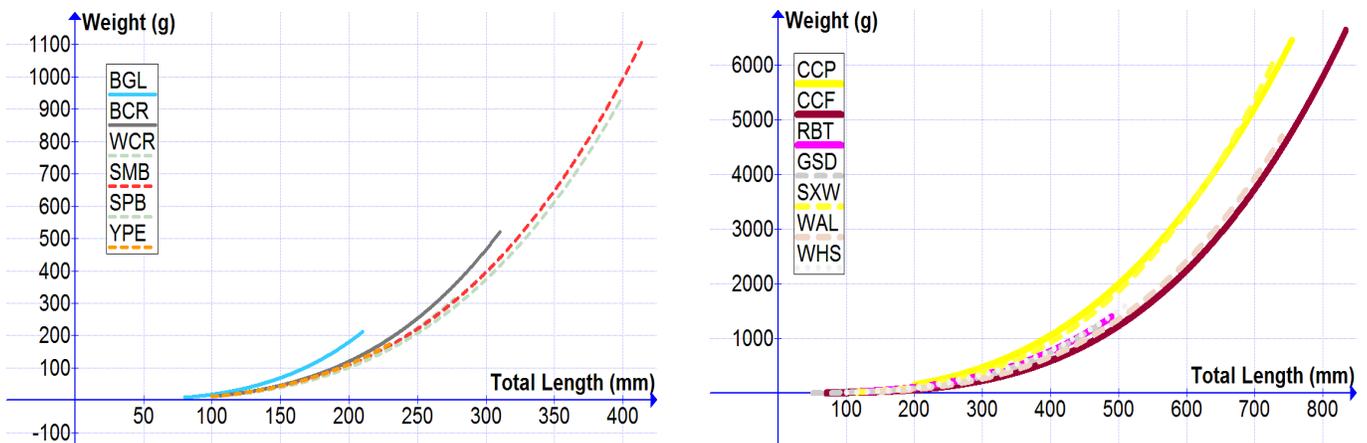

Figure 1: Graphs of standard weight equations for six smaller species studied (left) and seven larger species studied (right).

**Results**

The mean relative weights vs. year are shown in Table 2 for all the years where there is sufficient survey data. Some species such as channel catfish and spotted bass have a mean relative weight above 100 for most years showing that these species are consistently plump in Pueblo Reservoir. Other species such as walleye, rainbow trout, and common carp are have mean relative weights below 100 most years, with many years below 90 showing these species tend to be thin in Pueblo Reservoir. Some species such as smallmouth bass, gizzard shad, and hybrid striped bass show significant year to year variations in relative weight.

| Year | BGL | CCF | BCR | CPP | GSD | RBT | SMB | SPB | SXW | WAL | WCR | WHS | YPE |
|---|---|---|---|---|---|---|---|---|---|---|---|---|---|
| 1990 |  | 108.5 |  | 83.7 | 85.7 |  | 86.5 |  | 85.3 | 88.0 |  | 93.8 |  |
| 1992 |  | 100.9 |  | 95.7 | 86.4 | 95.7 | 84.9 |  | 92.3 | 91.5 | 89.9 | 97.5 |  |
| 1993 | 64.3 | 118.2 |  | 96.8 | 101.3 | 82.3 |  |  | 97.0 | 112.8 |  |  | 82.9 |
| 1994 | 98.7 | 118.9 |  | 89.9 | 85.8 | 92.2 | 100.7 | 115.6 | 94.6 | 88.6 | 99.8 | 95.4 | 88.8 |
| 1995 | 103.2 | 105.6 |  | 82.3 | 94.2 | 86.1 | 103.3 | 114.0 | 91.3 | 83.2 | 100.7 | 101.5 | 94.8 |
| 2001 |  | 101.3 |  | 81.0 | 90.7 | 83.5 | 84.6 | 98.8 | 86.4 | 83.2 | 83.0 | 94.5 |  |
| 2003 | 118.6 | 106.6 |  | 87.9 | 81.0 | 73.7 | 95.6 | 113.6 | 86.8 | 85.5 | 84.4 | 89.8 |  |
| 2004 |  | 111.2 | 102.8 | 85.7 | 106.4 | 87.7 | 92.3 | 108.1 | 96.0 | 89.3 |  | 102.9 |  |
| 2006 | 97.1 | 103.7 | 92.6 | 89.1 | 80.4 | 92.7 | 93.1 | 110.7 | 92.1 | 85.0 |  | 95.6 | 85.1 |
| 2007 | 110.6 | 104.1 | 104.3 | 83.0 | 85.2 |  | 104.0 | 108.4 | 88.6 | 84.8 | 86.1 | 96.2 | 95.7 |
| 2008 | 114.2 | 97.1 | 130.7 | 79.1 | 84.8 | 89.2 | 99.1 | 113.9 | 87.6 | 85.7 | 85.9 | 93.6 | 112.1 |
| 2009 | 102.2 | 102.5 |  | 88.3 | 90.8 | 80.0 | 90.1 | 106.2 | 93.6 | 91.9 | 90.4 | 102.1 |  |
| 2010 |  | 93.9 |  | 88.1 | 88.9 | 82.8 | 84.8 | 100.4 | 95.5 | 91.4 | 85.4 | 98.5 |  |
| 2011 | 113.7 | 96.3 | 95.5 | 85.3 | 83.4 |  | 82.4 | 98.9 | 92.8 | 86.3 |  | 99.4 |  |

*Table 2: Annual mean relative weights for different species in Pueblo Reservoir 1990 to 2011. The bottom row is the average Wr over available years.*

Figure 2 shows year to year relative weight vs. year for 13 species of fish in Pueblo reservoir from 1990 to 2011. Inspection of the graphs seem to suggest that some of the relative weights tend to increase together (positive correlation) and some relative weights tent to decrease when others increase (negative correlation). Computation of actual correlation coefficients (below) does a better job of quantifying these relationships.

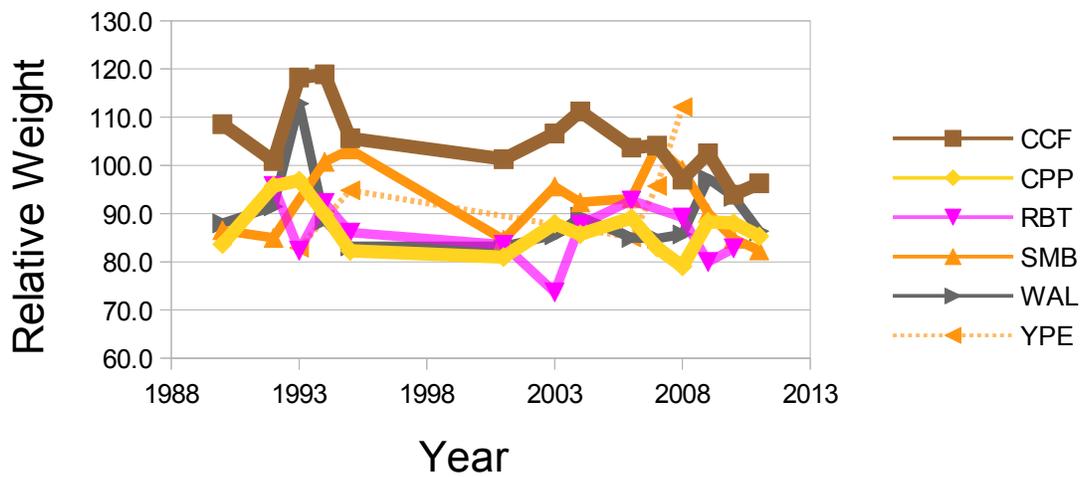
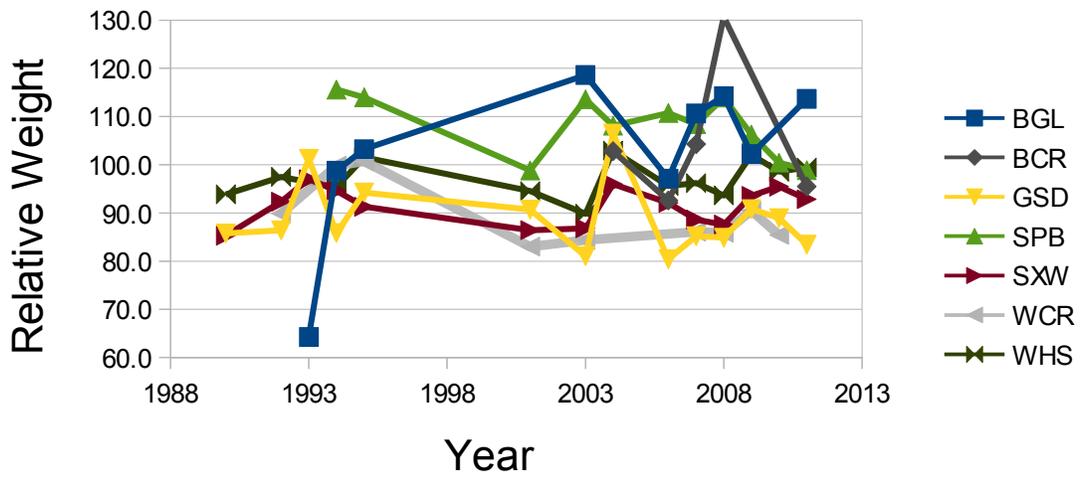

*Figure 2: Relative weight vs. year for species in Pueblo Reservoir from 1990 to 2011.*

*Bluegill*

Annual relative weights (Wr) and associated metrics for bluegill in Pueblo Reservoir are shown in Table 3 and Figure 3a. In 1993, the mean Wr of bluegill was 64.3 suggesting an overpopulation of bluegill relative to their food supply. One might think that the sample size of 3 might render the mean relative weight meaningless, but the notion of a large population of bluegill that year is also consistent with the largest or second largest relative weight for several of their predators: channel catfish, common carp, hybrid striped bass, and walleye as shown in Table 2. Common carp are not usually considered a bluegill predator, and actually are believed to be strong food competitors with bluegill in many ecosystems; however, the carp in Pueblo Reservoir are not very well fed and are often piscivorous. Anglers in Pueblo Reservoir report commonly catching carp on fish imitating crank baits and spoons.

| BGL | Wr | SEM | TL min | TL max | TL mean | n | SD |
|---|---|---|---|---|---|---|---|
| 1993 | 64.3 | 5.0 | 90 | 100 | 96.7 | 3 | 8.7 |
| 1994 | 98.7 | 4.5 | 80 | 180 | 122.7 | 92 | 43.4 |
| 1995 | 103.2 | 1.7 | 80 | 170 | 119.8 | 177 | 22.9 |
| 2003 | 118.6 | 5.3 | 80 | 190 | 113.0 | 47 | 36.2 |
| 2006 | 97.1 | 1.1 | 80 | 205 | 113.4 | 289 | 18.3 |
| 2007 | 110.6 | 4.0 | 80 | 190 | 110.1 | 35 | 23.5 |
| 2008 | 114.2 | 2.2 | 80 | 190 | 115.9 | 229 | 33.8 |
| 2009 | 102.2 | 6.6 | 130 | 185 | 163.3 | 3 | 11.5 |
| 2011 | 113.7 | 7.4 | 89 | 210 | 168.7 | 22 | 34.7 |

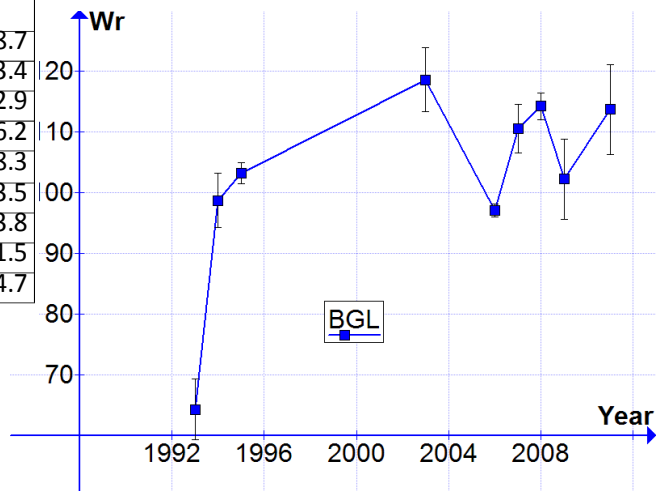

*Table 3/Figure 3a: Annual mean relative weights (Wr), standard error of the mean (SEM), minimum total length (TL min), maximum total length (TL max), and mean total length (TL mean) are shown along with the sample size (n), and the standard deviation from the mean of the relative weights (SD) for bluegill in Pueblo Reservoir.*

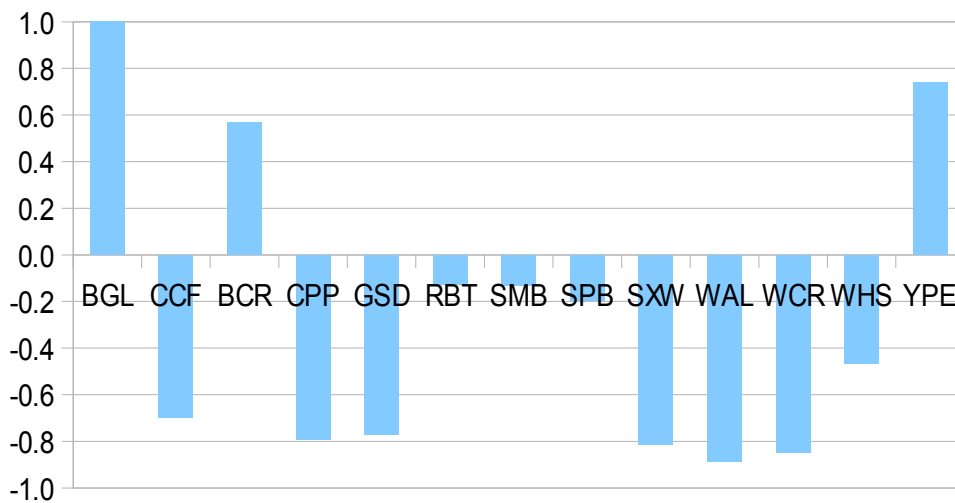

*Figure 3b: Correlation coefficients for annual mean relative weights between bluegill and other species in Pueblo Reservoir from 1993 to 2011. Note positive correlations with species with significant overlaps in forage habits (black crappie and yellow perch) and negative correlations with predators channel catfish, common carp, hybrid striped bass, and walleye.*

Figure 3b shows the correlations of annual relative weight of other species with bluegill. As expected, the correlations are positive with black crappie and yellow perch, the main food competitors and the correlations are negative with species which prey significantly on bluegill. It is notable that competition with yellow perch and predation by walleye are both significant in a much more complex food web just as they were found to significant in a system that included only walleye, yellow perch, and bluegill (Schneider, 1997). Addition of bluegill to a lake previously inhabited by only yellow perch and walleye increased the total fish biomass by 78%, increased the walleye biomass by 11%, and decreased the yellow perch biomass by 20%. The abundance of large walleye increased by 54% and the abundance of large yellow perch decreased by 76%. The increase in walleye production was attributed to utilization of bluegill as food. The decrease in yellow perch production was attributed to competition with bluegill for large zooplankton and benthos.

Several features of Figure 3b were surprising at first: the positive correlation of Wr with black crappie seemed reasonable, but not along with the strong negative correlation with white crappie. If black crappie and white crappie both have the same feeding habits and occupy similar niches in the food web dynamics, one would expect a similar correlation with bluegill. However, previous studies have found that black crappie diet was dominated almost exclusively by zooplankton and insects, and that this was also true for white crappie up to 200 mm long (Ellison, 1984). Black crappie longer than 200 mm did not transition as readily to eating fish as white crappie. The black crappie in the study had average lengths that ranged from 133 mm to 235 mm, depending on year. The white crappie averaged from 179 mm to 323 mm. The greater lengths and stronger piscivory of the white crappie explain the strong negative correlation of the white crappie Wr with that of bluegill. The shorter lenghts and focus on zooplankton and other invertebrates explains the strong positive correlation of black crappie Wr with bluegill.

The strong negative correlation of bluegill Wr with gizzard shad Wr is surprising. One wonders if high Wr in adult gizzard shad produces a high level of fecundity leading to a strong age zero cohort of gizzard shad which then competes strongly with the bluegill.

Rainbow trout, smallmouth bass, and spotted bass all have a small negative correlation of Wr with bluegill Wr. One might reasonably suppose either there simply might not be significant overlaps in the occupied niches or that both strong predatory and strong competition relationships exist between bluegill and these other species. Since the mean TL of smallmouth bass in the study ranged from 173 mm to 278 mm, it is reasonable to suggest that both a competitive relationship (mostly with the smaller smallmouth bass) and a predatory relationship (mostly with the larger smallmouth bass) exists between the two species. There was ample data for smallmouth bass to test this suggestion by computing relative weights for different length classes of smallmouth bass. The annual Wr of smallmouth bass between 150 mm and 180 mm in total length had a strong positive correlation (0.607) with bluegill, which probably reflects significant overlap of food supply with bluegill, because it indicates that they both tend to be plumper in years where there is plenty of food, and they both tend to be thinner in years when food is scarce. In contrast, the annual Wr of smallmouth bass between 280 mm and 430 mm total length showed a negative correlation (-0.303) with the annual Wr of bluegill. This is consistent with the suggestion that smallmouth bass in this length range are more piscivorous and tend to be fatter when bluegill are so well populated that the bluegill tend to be thinner due to intraspecific competition.

Likewise, since the mean TL of spotted bass in the study ranged from 158 mm to 321 mm, both competitive and predatory relationships with bluegill are suggested here also. One might expect that the longer length classes of spotted bass have a more strongly negative correlation of their Wr with bluegill; whereas, the shorter length classes of spotted bass would likely have a positive correlation of their Wr with bluegill. However, there is not sufficient data on the spotted bass to analyze Wr correlations for separate length classes. The small negative correlation with rainbow trout is hard to explain, as few of the rainbow trout in the sample are over 400 mm long, their feeding habits suggest more competition than predation with bluegill. It may be that rainbow trout are not actually eating many bluegill above 80 mm total length (the minimum for computing Wr), but that the smaller bluegill that they are eating have such strong food overlap with the bluegill included in Wr computations, that there is a negative correlation of Wr between bluegill and rainbow trout.

*Channel Catfish*

Mean annual relative weights (Wr) and associated metrics for channel catfish in Pueblo Reservoir are shown in Table 4 and Figure 4a. The mean relative weights above 100 most years, are higher than most other species in the reservoir. Channel catfish are thriving in Pueblo Reservoir, making excellent use of resources, and competing well with other species. With total lengths up to 834 mm, channel catfish are the largest predator present in the reservoir in significant numbers. The mean relative weight of channel catfish has trended slightly downward in the last few years. This may be due to increased stocking of channel catfish due to their success in the reservoir and the popularity of the species among anglers or it may be due to the decline of the yellow perch in the reservoir.

The channel catfish is one of the few piscivorous species in Pueblo Reservoir showing a low correlation of Wr with total length ($R^2 = 0.055$). In channel catfish, relative weights tend to increase very slightly with total length, about 2.1 for every additional 100mm of length. Most piscivores in the reservoir show decreasing Wr with length class. Maintaining Wr with longer lengths may be due to lower metabolic needs or a broader forage base. In fat years, channel catfish are capable predators able to competing well with other predators. In lean years, channel catfish are also capable scavengers able to maintain body condition by eating the remains of other species.

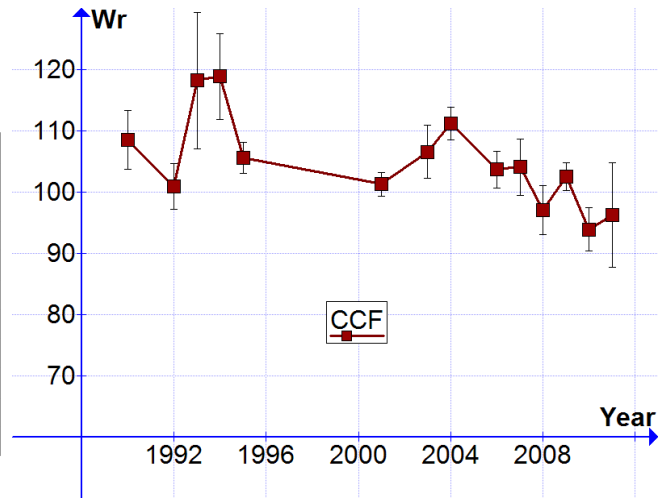

| CCF | Wr | SEM | TL min | TL max | TL mean | n | SD |
|---|---|---|---|---|---|---|---|
| 1990 | 108.5 | 4.8 | 355 | 680 | 521.2 | 13 | 17.2 |
| 1992 | 100.9 | 3.7 | 225 | 620 | 486.0 | 5 | 8.2 |
| 1993 | 118.2 | 11.1 | 332 | 727 | 447.0 | 4 | 22.1 |
| 1994 | 118.9 | 7.0 | 205 | 700 | 419.6 | 12 | 24.2 |
| 1995 | 105.6 | 2.5 | 230 | 625 | 391.5 | 10 | 7.9 |
| 2001 | 101.3 | 1.9 | 280 | 800 | 490.7 | 22 | 9.0 |
| 2003 | 106.6 | 4.3 | 430 | 670 | 551.7 | 6 | 10.6 |
| 2004 | 111.2 | 2.7 | 330 | 800 | 581.7 | 18 | 11.4 |
| 2006 | 103.7 | 3.0 | 340 | 720 | 510.8 | 12 | 10.4 |
| 2007 | 104.1 | 4.6 | 350 | 770 | 613.6 | 11 | 15.1 |

*Table 4/Figure 4a: Annual mean relative weights, standard error of the mean, minimum total length, maximum total length, and mean total length are shown along with the sample size, and the standard deviation from the mean of the relative weights for channel catfish in Pueblo Reservoir.*

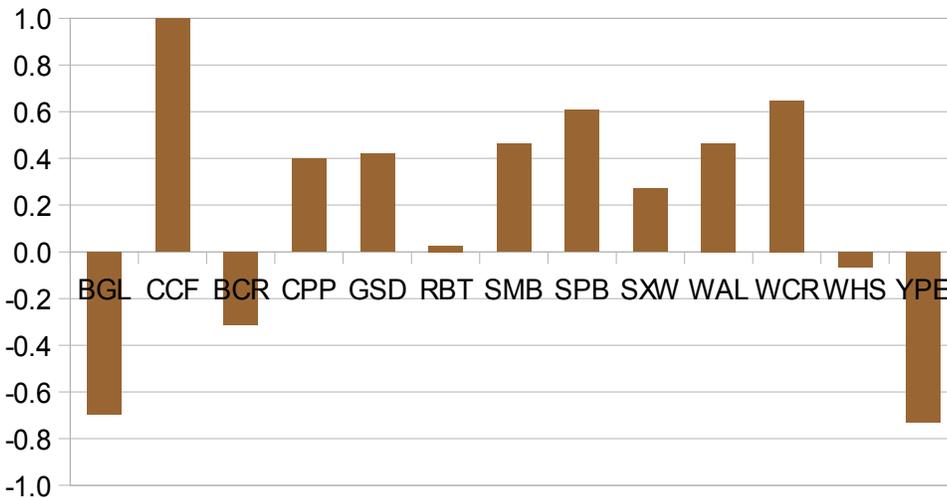

*Figure 4b: Correlation coefficients for annual mean relative weights between channel catfish and other species in Pueblo Reservoir from 1990 to 2011. Note positive correlations with species with significant overlaps in forage habits and negative correlations with prey species including bluegill, black crappie, and yellow perch.*

Figure 4b shows the correlations of annual relative weight of other species with channel catfish. As expected, the correlations are positive with species with significant forage overlaps with the channel catfish, and the correlations are strongly negative with bluegill and yellow perch, important prey species and slightly negative with black crappie, another important prey species.

Significant correlations of channel catfish relative weight with relative weight of common carp (r = 0.399) and gizzard shad (r = 0.420) were unexpected, because these species are not believed to have significant forage overlap with channel catfish. However, the fecundity of adult gizzard shad and quality index of age zero gizzard shad have been shown to be well correlated with the relative weight of the adult cohort of gizzard shad (Willis, 1987), so it seems reasonable to hypothesize that the correlation of Wr between channel catfish and adult gizzard shad is due to a strong age 0 class of gizzard shad upon which the channel catfish are feeding in years where gizzard shad have a high Wr. The fecundity hypothesis may also explain the correlation with relative weight of common carp and is perhaps further supported by observing that Wr of channel catfish is more strongly correlated with Wr of common carp from the past year (r = 0.694). This suggests that the cohort of age zero common carp available for channel catfish to eat may strongly depend on the Wr of common carp going into the previous fall. Blackwell et al. (2000) summarize the relationship of relative weight with fecundity and reproductive potential.

The correlation of channel catfish Wr with Wr of smallmouth bass is 0.461, indicating a significant overlap in forage. Figure 4c shows the correlation of channel catfish with different length classes of smallmouth bass (Anderson and Neumann, 1996). There is a positive correlation with all lengths of smallmouth bass, but there is a higher correlation (r = 0.648) for the quality and preferred combined length classes of smallmouth bass, indicating the likelihood of greater forage competition between channel catfish and the longer smallmouth bass with total lengths greater than 280 mm.

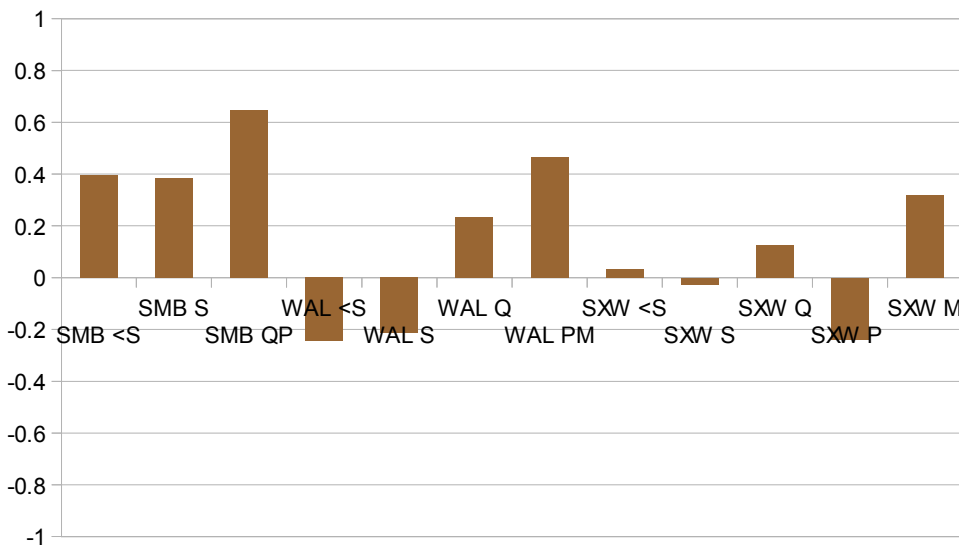

*Figure 4c: Correlation coefficients for annual mean relative weights between channel catfish and available length classes of smallmouth bass, walleye, and hybrid striped bass. The length class shorter than stock is < S. Other length classes are designated by S (stock), Q (quality), P (preferred), and M (memorable). Length classes may be combined if necessary for analysis.*

Figure 4c shows that correlation of Wr increases between channel catfish and walleye as the length class of walleye increases. Walleye below stock length (150-250 mm) have a correlation coefficient of -0.245. This suggests the possibility of predation of channel catfish on this length class of walleye, which is not otherwise known or mentioned in the literature (to the authors' knowledge). However, this is not unreasonable, since channel catfish are known to prey on yellow perch of comparable lengths (authors' personal experience), and walleye at the lower end of the length range (150 mm) have been found in stomach contents of northern pike (mean TL 500 mm) and walleye prey of 100 mm TL have been found in stomachs of smallmouth bass and walleye predators near 300 mm TL (Liao et al., 2002) . However, the negative correlation with stock length (250-380 mm) walleye is harder (but not impossible, given the top lengths of channel catfish) to interpret as due to predation, and may be due to scavenging of carcasses that died of other causes or some other factor. Starvation and hunger related mortality probably increases for walleye with Wr < 80, and in years with a mean Wr in the low 80s, there were a significant number of walleyes with Wr < 80. Another possibility is that channel catfish might be preying on other scavengers feeding on carcasses of walleye in this length class (crawfish, etc.) Walleye in the quality and in the preferred and memorable combined length classes have the expected positive correlations suggesting a significant overlap in forage.

Channel catfish have no significant correlation of their Wr with hybrid striped bass in and below the stock length class. The correlation with hybrid striped bass in the quality length range is slightly positive (r = 0.126), but the correlation with hybrid striped bass in the preferred length class (380-510 mm) is negative (r = -0.239) even though common predation on bluegill, shad, crawfish, crappie, and yellow perch suggests that it should be positive. However, this length class has significant numbers of fish with Wr below 80, so it is possible that starvation and other hunger related mortality are producing an increase in scavenging forage for channel catfish in leaner years. The relative weight of hybrid striped bass in the memorable length class (> 510 mm) is positively correlated with the mean annual relative weight of channel catfish, as would be expected from the common prey species.

*Walleye*
Annual relative weights (Wr) and associated metrics for walleye in Pueblo Reservoir are shown in Table 5 and Figure 5a. The mean Wr in the reservoir is 89.1, suggesting walleye tend to be thin.

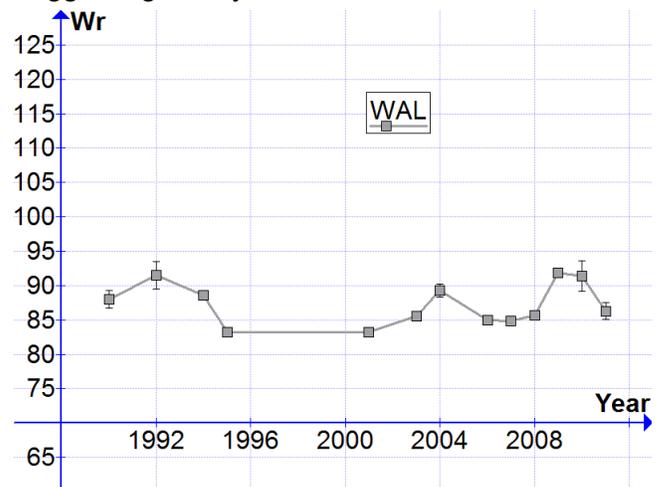

*Table 5a/Figure 5a: Annual relative weights, standard error of the mean, minimum total length, maximum total length, and mean total length are shown along with the sample size, and the standard deviation from the mean of the relative weights for walleye in Pueblo Reservoir.*

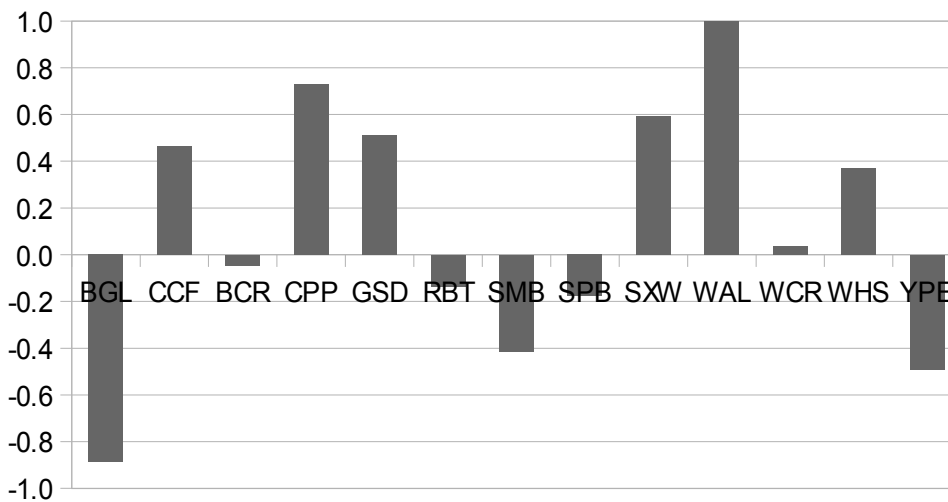

*Figure 5b: Correlation coefficients for annual mean relative weights between walleye and other species in Pueblo Reservoir from 1990 to 2011. Note positive correlations with species with significant overlaps in forage habits and negative correlations with prey species including bluegill, and yellow perch.*

Figure 5b shows the correlation between the annual Wr for walleye and the Wr for other species. The correlation is strongly negative with bluegill, suggesting that bluegill are an important prey species. In years where there are so many bluegill that intraspecific competition causes bluegill Wr to be lower, walleye Wr tends to be higher, because bluegill prey are plentiful. Yellow perch are also an important prey species for walleye; consequently, they have a significant negative correlation with walleye.

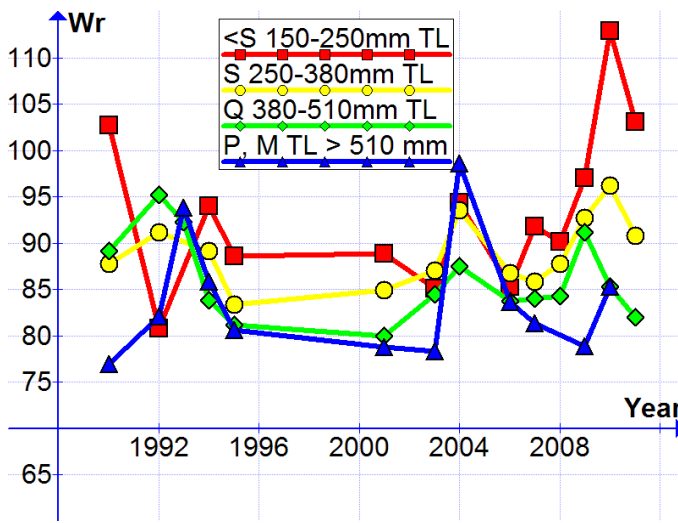

*Figure 5c: Annual mean Wr for different length classes of walleye. In most years, the mean Wr of walleye are rank ordered, with the shortest walleye (TL 150 mm to 250 mm) having the highest Wr, and the longest (TL > 510 mm) having the smallest Wr. The most prominent exceptions are 1993 and 2004, years when bluegill and black crappie stockings provided a significant influx of forage and a boost to the relative weights of the larger walleye. The mean annual relative weights (all years) are below stock length (<S), 93.4; stock length class (S), 89.0; quality length class (Q), 86.0; preferred and memorable combined length classes (P, M), 84.5.*

Porath and Peters (1997) showed that studying relative weights of different length classes of walleye can be a cost effective way to assess prey availability. Figure 5c shows the annual mean relative weights for different length classes of walleye found in Pueblo Reservoir. The survey sample sizes are sufficient most years to compute mean relative weights in different length classes. This allows for computation of correlations between Wr of different length classes of walleye with different species of fish to identify potential interspecific interactions with different length classes of walleye.

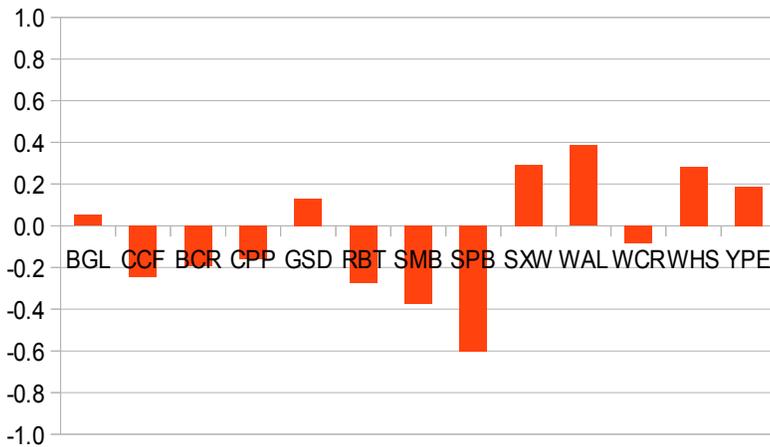

*Figure 5d: Correlations between annual Wr of other species and Wr of the length class of walleye from 150 mm to 250 mm TL. Note slight positive correlations with bluegill, hybrid striped bass, walleye (all length classes), and yellow perch, suggesting competition and forage overlap. Positive correlation with gizzard shad and white sucker is hypothesized to be due to high fertility in those species in high Wr years resulting in strong crops of age zero gizzard shad and white suckers becoming forage for these walleye.*

Figure 5d shows the correlation of Wr of different species with the shortest length class of walleye, those below stock length of 250 mm, but above the minimum 150 mm for computing relative weights. Significant negative correlations with smallmouth bass and spotted bass are hard to understand. Perhaps high Wr in adults of these species causes high reproductive potential and strong age zero cohorts, which then compete strongly for available forage with walleye in this length class.

Figure 5e shows the correlation of Wr of different species with the stock length class of walleye (250 mm < TL < 380 mm). Significant negative correlations with smallmouth bass and spotted bass are hard to understand. It is possible that high Wr in adults of these species causes high reproductive potential and strong age zero cohorts, which then compete strongly for available forage with walleye in this length class. In any case, negative correlations in Wr between walleye and smallmouth bass is consistent with the results of Wuellner et al. (2011) in four of six South Dakota lakes. Positive correlation with common carp, gizzard shad, and white sucker is hypothesized to be due to high fertility in those species in high Wr years resulting in strong crops of age zero offspring for walleye to eat. Perhaps the near zero correlation in yellow perch is due to nearly offsetting opposite interactions. Yellow perch are known to be an important prey species, so a negative correlation is expected attributable to yellow perch having a small Wr in years when they are abundant and intraspecific competition reduces their available food. However, it may also be that in Pueblo Reservoir, yellow perch also have significant interspecific competition with walleye in this length class. Interpecific competition without predation would suggest a positive correlation.

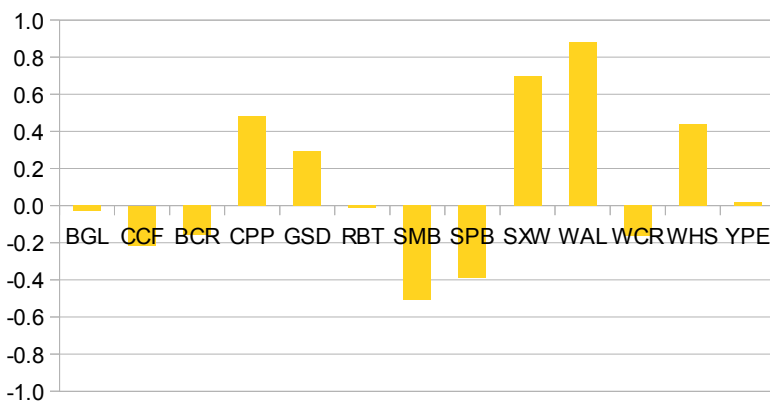

*Figure 5e: Correlations between annual Wr of other species and Wr of the length class of walleye from 250 mm to 380 mm TL. Note significant positive correlations with hybrid striped bass and walleye (all lengths), suggesting forage overlap.*

Figure 5f shows the correlation of Wr of different species with the quality length class of walleye (380 mm < TL < 510 mm). Note significant positive correlations with hybrid striped bass and walleye (all lengths), suggesting forage overlap. Note also that this is the smallest length class of walleye showing a significant negative correlation with bluegill and yellow perch, suggesting significant predation of these species. And as the correlation with these prey species becomes significantly negative, the correlation with channel catfish becomes positive, suggesting overlap in prey species. Once again, positive correlation with common carp, gizzard shad, and white sucker may be attributable to the hypothesis of increased age zero yield of these species with high Wr.

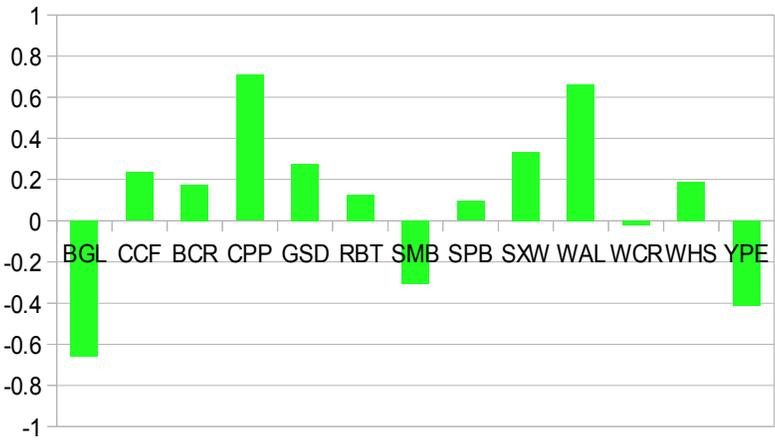

*Figure 5f: Correlations between annual Wr of other species and Wr of the length class of walleye from 380 mm to 510 mm TL. Note that the correlation with Wr of common carp has become increasingly positive as the length of walleye increases.*

Figure 5g shows the correlation of Wr of different species with the preferred and memorable combined length classes of walleye (TL > 510 mm). The negative correlations with bluegill and yellow perch are larger than in the smaller length classes of walleye. Likewise, the positive correlations with channel catfish and hybrid striped bass are also larger than with smaller length classes of walleye suggesting even greater forage overlap with these species, as the longer walleye become more strongly piscivorous. Significant positive correlations with common carp, gizzard shad, and white sucker may be attributable to the fecundity hypothesis.

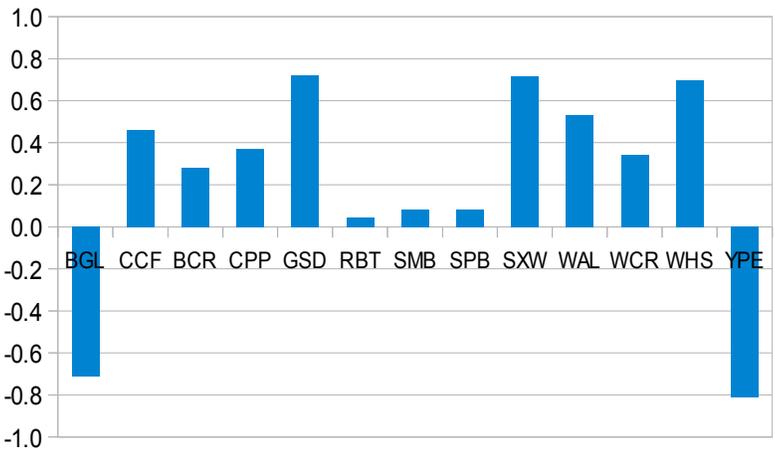

*Figure 5g: Correlations between annual Wr of other species and Wr of the length class of walleye with TL > 510 mm.*

|     | WAL <S | WAL S  | WAL Q  | WAL P, M |
|-----|--------|--------|--------|----------|
| BGL | 0.052  | -0.028 | -0.662 | -0.711   |
| CCF | -0.245 | -0.214 | 0.235  | 0.463    |
| BCR | -0.193 | -0.158 | 0.175  | 0.283    |
| CPP | -0.158 | 0.480  | 0.709  | 0.369    |
| GSD | 0.129  | 0.293  | 0.276  | 0.720    |
| RBT | -0.277 | -0.013 | 0.126  | 0.043    |
| SMB | -0.375 | -0.509 | -0.306 | 0.080    |
| SPB | -0.606 | -0.391 | 0.097  | 0.081    |
| SXW | 0.291  | 0.696  | 0.336  | 0.719    |
| WAL | 0.385  | 0.884  | 0.663  | 0.532    |
| WCR | -0.082 | -0.162 | -0.021 | 0.344    |
| WHS | 0.285  | 0.441  | 0.190  | 0.698    |
| YPE | 0.188  | 0.018  | -0.411 | -0.813   |

*Table 5b (left)/Figure 5h (below): Correlations between annual Wr of other species and Wr of different length classes of walleye. Note the correlation with bluegill and yellow perch decrease (become increasingly more negative) as the length class of walleye increases. Conversely, the correlation with channel catfish, black crappie, and gizzard shad become more positive as the length class of walleye increases. Correlations with common carp, gizzard shad, rainbow trout, spotted bass, hybrid striped bass, white crappie, and white sucker show an increasing, though not monotonic trend as the length class of walleye increases.*

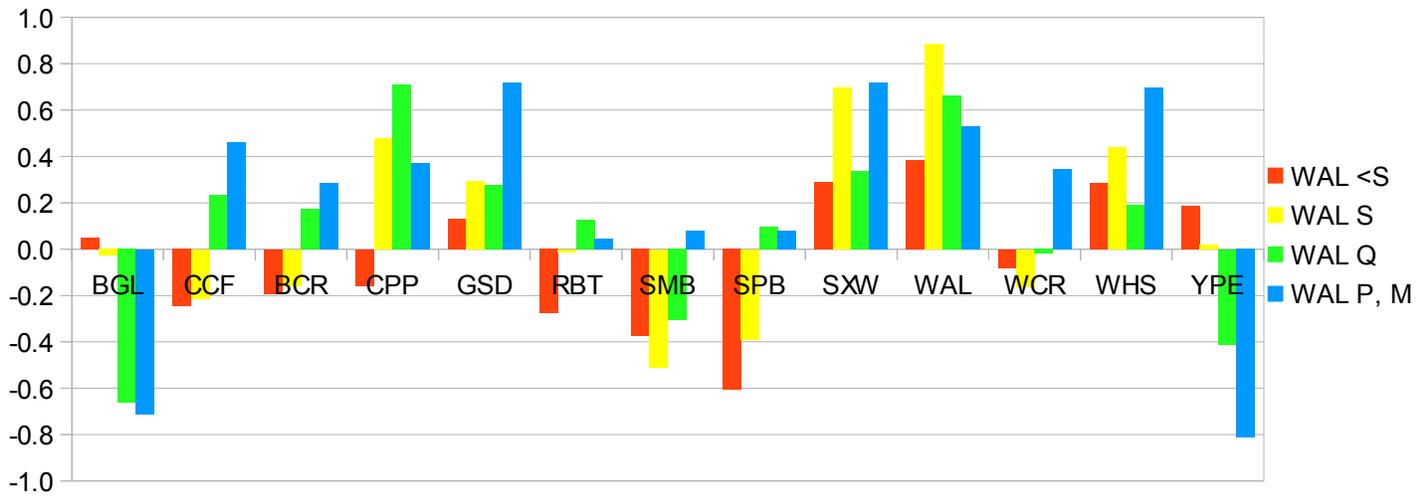

|         | SMB <S | SMB S  | SMB QP | WAL <S | WAL S | WAL Q  | WAL P, M | SXW <S | SXW S  | SXW Q | SXW P | SXW M |
|---------|--------|--------|--------|--------|-------|--------|----------|--------|--------|-------|-------|-------|
| WAL < S | -0.724 | -0.380 | -0.038 | 1.000  | 0.581 | -0.068 | 0.005    | -0.773 | 0.539  | 0.174 | 0.306 | 0.245 |
| WAL S   | -0.750 | -0.297 | -0.104 | 0.581  | 1.000 | 0.516  | 0.482    | -0.547 | 0.785  | 0.698 | 0.677 | 0.465 |
| WAL Q   | -0.111 | -0.196 | -0.018 | -0.068 | 0.516 | 1.000  | 0.426    | -0.683 | 0.634  | 0.418 | 0.259 | 0.129 |
| WAL P, M | -0.216 | 0.161 | 0.254  | 0.005  | 0.482 | 0.426  | 1.000    | 0.513  | 0.283  | 0.812 | 0.634 | 0.784 |

*Table 5c (above): Correlations between Wr of different length classes of walleye with different length classes of smallmouth bass and hybrid striped bass.*

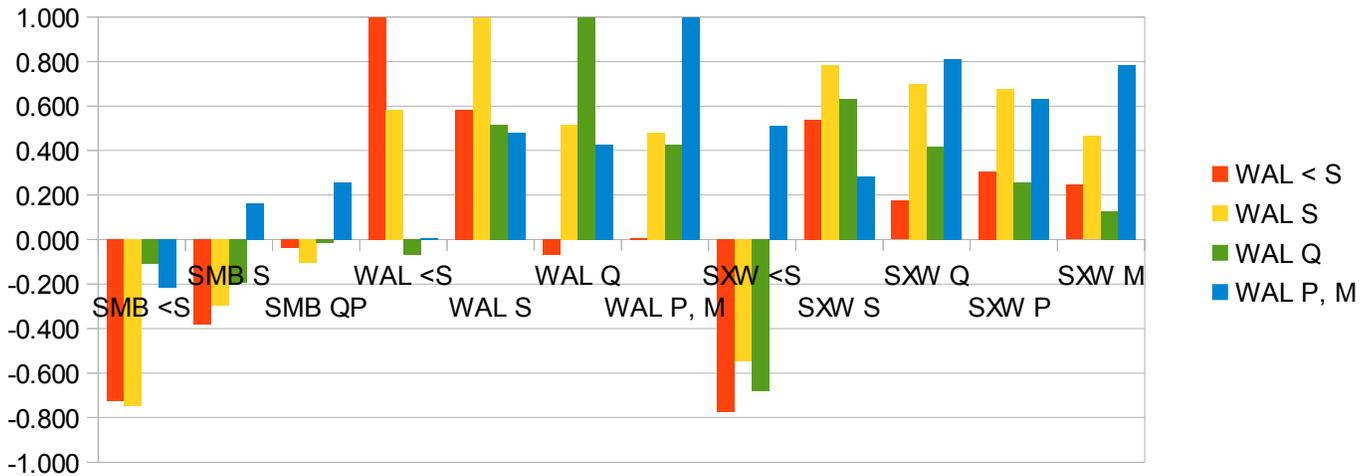

*Figure 5i: Correlations between Wr of different length classes of walleye with different length classes of smallmouth bass and hybrid striped bass. Note no significant positive correlations with smallmouth bass, suggesting a lack of competition. There are a number of significant positive correlations with other length classes of walleye and length classes of hybrid striped bass of stock length and longer suggesting significant competition.*

Current Colorado Division of Wildlife practice has been to stock significant quantities of walleye every year in Pueblo Reservoir. The present study has shown that the relative weights of walleye are below optimal and decrease with increasing length class. This suggests that the population of walleye is too high relative to their available food supply which limits both growth and body condition and may also reduce natural reproduction in the reservoir (Kohler and Kelly, 1991; Blackwell et al., 2000; Brown et al., 1990). Systems such as the Eleven Point River, Arkansas, (Henry et al., 2008) Hauser Reservoir, Montana, and Holter Reservoir, Montana (Roberts and Dalbey, 2007) where stocked walleye have high relative weights which are level or slightly increasing with walleye length class demonstrate evidence predator-prey balance. In contrast, walleye at Pueblo Reservoir demonstrate high intraspecific competition and low relative weights more like Sylvan Lake and Winona Lake, Indiana (Burlingame et al., 2006). Figure 5i and Table 5c show significant levels of interspecific competition between difffferent length classes of walleye and different length classes of hybrid striped bass. This data also demonstrates significant intraspecific competition between walleye length classes at Pueblo Reservoir. Since hybrid striped bass populations are maintained through stocking and walleye populations receive significant augmentation through stocking, body condition and growth rates of walleye would likely be significantly increased by a reduction in stocking densities of walleye and hybrid striped bass.

*Yellow Perch*

Annual relative weights (Wr) and associated metrics for yellow perch in Pueblo Reservoir are shown in Table 6 and Figure 6a. Yellow perch have not been stocked in the reservoir in several years, and their survey numbers have declined significantly, with only one fish in the survey in 2010 and none in 2011. When present in significant numbers, yellow perch are an important prey species for walleye, hybrid striped bass, and channel catfish. However, yellow perch function as an invasive species with a number of negative impacts when and if introduced to nearby reservoirs managed for cool water species, and illegal stocking of yellow perch is an ongoing problem among higher altitude reservoirs in Colorado. Note that as yellow perch numbers dwindled from 2004 to 2008, their mean relative weight has increased, probably due to reduced intraspecific competition for food. But over the same period, Wr of channel catfish, hybrid striped bass, and walleye, main predators, tended to decrease. However, note that since bluegill mean annual Wr was also increasing over the same period, the decrease in Wr among these three predators may be due to a combined effect of reduced numbers of both prey species and not reduced numbers of yellow perch alone.

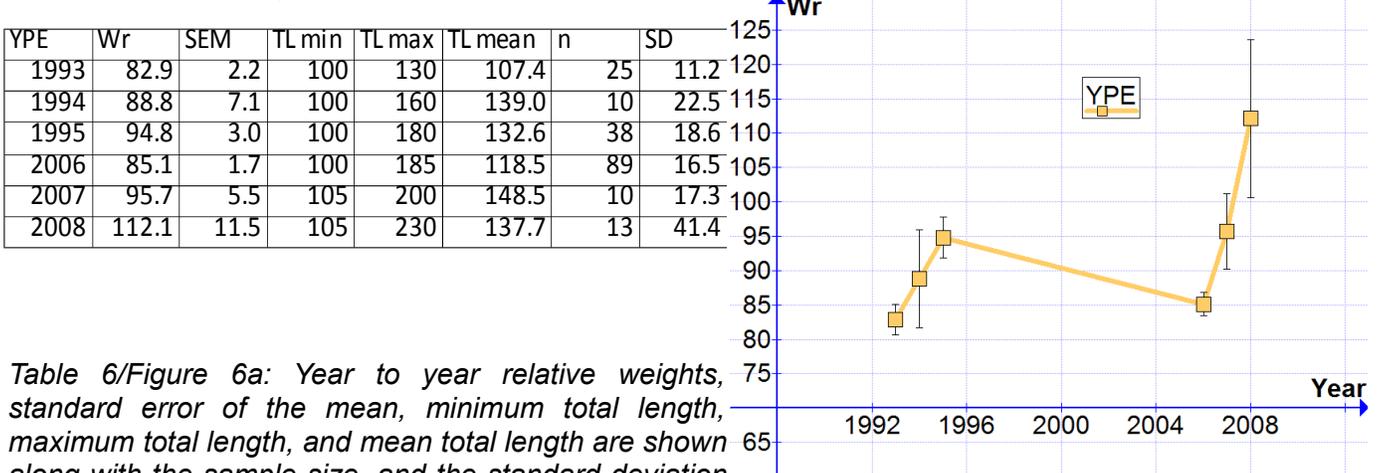

| YPE | Wr | SEM | TL min | TL max | TL mean | n | SD |
|---|---|---|---|---|---|---|---|
| 1993 | 82.9 | 2.2 | 100 | 130 | 107.4 | 25 | 11.2 |
| 1994 | 88.8 | 7.1 | 100 | 160 | 139.0 | 10 | 22.5 |
| 1995 | 94.8 | 3.0 | 100 | 180 | 132.6 | 38 | 18.6 |
| 2006 | 85.1 | 1.7 | 100 | 185 | 118.5 | 89 | 16.5 |
| 2007 | 95.7 | 5.5 | 105 | 200 | 148.5 | 10 | 17.3 |
| 2008 | 112.1 | 11.5 | 105 | 230 | 137.7 | 13 | 41.4 |

*Table 6/Figure 6a: Year to year relative weights, standard error of the mean, minimum total length, maximum total length, and mean total length are shown along with the sample size, and the standard deviation from the mean of the relative weights for yellow perch in Pueblo Reservoir.*

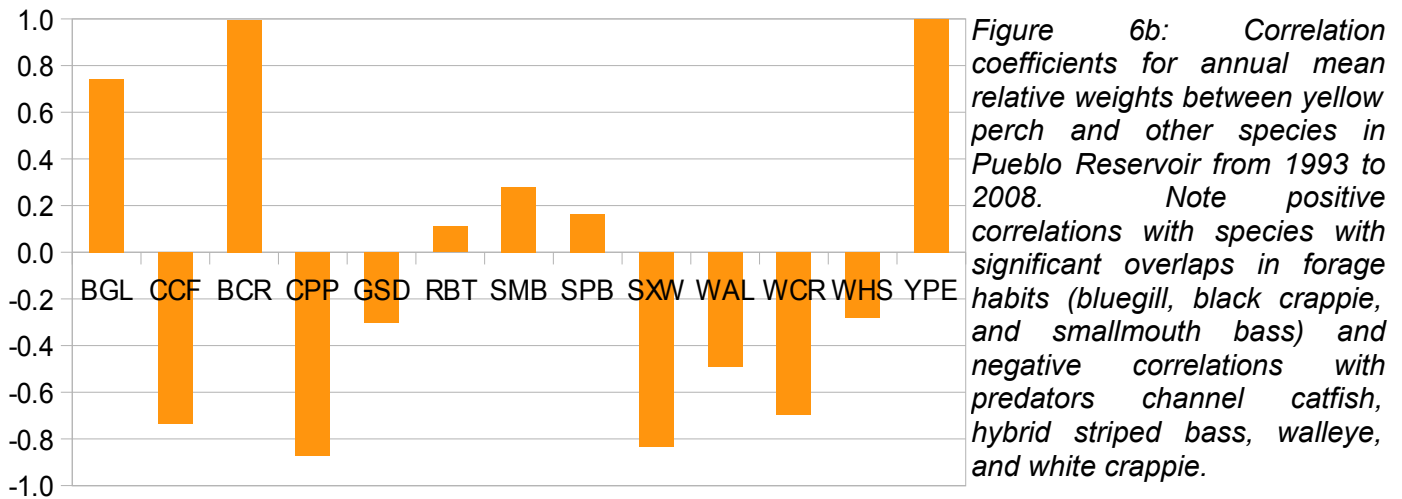

*Figure 6b: Correlation coefficients for annual mean relative weights between yellow perch and other species in Pueblo Reservoir from 1993 to 2008. Note positive correlations with species with significant overlaps in forage habits (bluegill, black crappie, and smallmouth bass) and negative correlations with predators channel catfish, hybrid striped bass, walleye, and white crappie.*

Figure 6b shows the correlations of annual relative weight of other species with yellow perch. As expected, the correlations are positive with black crappie and bluegill, the main food competitors. Correlations are negative with species which prey significantly on yellow perch, including channel catfish, common carp, hybrid striped bass, walleye, and white crappie. Yellow perch are well known as an important prey species in many ecosystems, including complex food webs with many species present such as Spirit Lake, Iowa (Liao et al., 2002).

Some readers may be surprised by the identification of common carp as a predator on yellow perch.  The authors of the present study have been unable to find documented examples of common carp predation on yellow perch in the literature either through stomach content analysis or isotope analysis.  Yet, the Wr of common carp is the most negatively correlated of the species studied here.  There may be a contribution to this negative correlation by high Wr years in common carp leading to strong age zero classes of common carp which have the effect of reducing yellow perch Wr via interspecific competition.  However, the three year pattern of increasing yellow perch Wr from 1993 to 1995 and 2006 to 2008 suggest a large influx of yellow perch in 1993 and 2006 (either stocking or strong recruitment years) followed by a decline in population due to predation.  This suggests intraspecific competition among yellow perch rather than interspecific competition with age zero carp as the cause of low mean Wr for yellow perch.  Therefore, the hypothesis of significant predation of common carp on yellow perch is favored.  The proportional stock density of surveyed common carp is 94, indicating that 94% of the common carp above minimum stock length (280 mm) are also above minimum quality length (410 mm).  In addition, 68% of the common carp above the minimum stock length are also above the minimum preferred length (530 mm).  Clearly, there are a lot of big carp in the reservoir.  Anglers report that carp are readily caught on crankbaits (such as popular Rapala plugs) either trolled or cast from shore, and some anglers even complain that they catch too many carp when targeting walleye trolling crankbaits and spoons.

    It is no surprise that analysis of Wr correlation suggests that channel catfish, walleye, and hybrid striped bass are significant predators on yellow perch.  The suggestion that white crappie might be preying on yellow perch longer than 100 mm (the minimum TL for computing Wr) is more interesting.  However, most years show a number of white crappie over 300 mm TL, and there is also the possibility that the Wr of yellow perch above 100 mm is highly correlated with the condition of shorter yellow perch, even though the shorter yellow perch are excluded from Wr calculations.

    Yellow perch are believed to be an important part of walleye diets in ecosystems where the two species occur together in abundance.  However, there are few studies indicating the length classes of walleye where predation is important.  Figure 6c shows the correlation of yellow perch Wr with different length classes of walleye (Anderson and Neumann, 1996).  Yellow perch seem to be food competitors rather than prey with walleye below stock length (250 mm).  Effects of competition and predation seem to be canceling out to yield a correlation close to zero with stock length walleye (250mm to 380 mm).  Predation on yellow perch by quality length walleye appears to become stronger and produces a negative correlation  (380 mm to 510 mm).  Predation of yellow perch by preferred and memorable length walleye (TL > 510 mm) appears to be the strongest as indicated by a Wr correlation of r = -0.813.

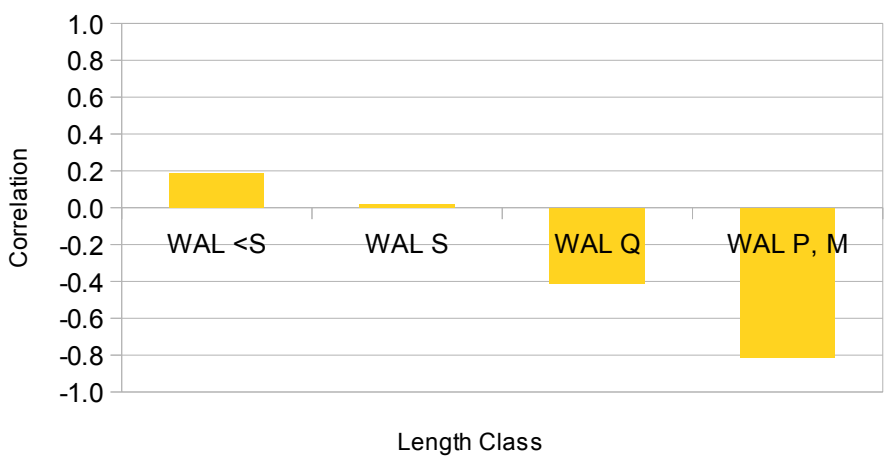

*Figure 6c: Correlation coefficients for annual mean relative weights between yellow perch and walleye by length class in Pueblo Reservoir from 1990 to 2011.*

*Smallmouth Bass*

Smallmouth bass are a popular sport fish, and the Pueblo Reservoir bass fishery has a dedicated group of anglers. Annual relative weights (Wr) and associated metrics for smallmouth bass in Pueblo Reservoir are shown in Table 7a and Figure 7a. The mean relative weight over the time period is 93.0. In plump years, the mean Wr has been as high as 104; whereas, in lean years, the mean Wr has dipped as low as 82.4. The mean relative weight of smallmouth bass below stock size (150 mm to 180 mm) is 100.8 suggesting there is plenty of available forage for this length class. The mean relative weight for stock length smallmouth bass (180 mm to 280 mm) is 93.6, and the mean relative weight for the quality and preferred combined length classes is 85.9, suggesting that as they get longer and shift to greater piscivory, smallmouth bass body condition is limited by available forage.

| SMB  | Wr    | SEM | TL min | TL max | TL mean | n   | SD   |
|------|-------|-----|--------|--------|---------|-----|------|
| 1990 | 86.5  | 5.9 | 200    | 255    | 172.7   | 2   | 8.3  |
| 1994 | 100.7 | 1.1 | 150    | 375    | 215.7   | 241 | 17.6 |
| 1995 | 103.3 | 1.5 | 150    | 380    | 205.3   | 105 | 15.8 |
| 2001 | 84.6  | 1.4 | 200    | 340    | 252.2   | 29  | 7.6  |
| 2003 | 95.6  | 1.6 | 150    | 360    | 221.3   | 92  | 15.2 |
| 2004 | 92.3  | 2.9 | 210    | 300    | 248.0   | 5   | 6.6  |
| 2006 | 93.1  | 1.1 | 150    | 400    | 194.9   | 129 | 12.2 |
| 2007 | 104.0 | 1.7 | 150    | 400    | 203.9   | 152 | 20.4 |
| 2008 | 99.1  | 1.1 | 150    | 415    | 210.2   | 193 | 15.6 |
| 2009 | 90.1  | 2.3 | 150    | 410    | 265.6   | 17  | 9.6  |
| 2010 | 84.8  | 1.2 | 176    | 381    | 277.7   | 20  | 5.2  |
| 2011 | 82.4  | 1.7 | 204    | 361    | 260.5   | 33  | 9.8  |

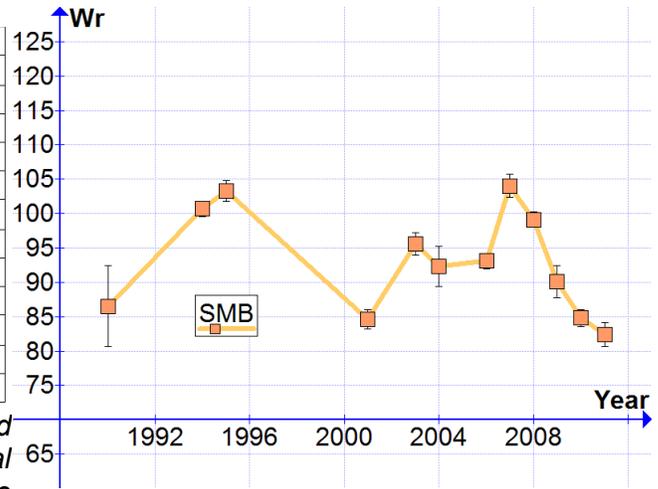

*Table 7a/Figure 7a: Annual relative weights, standard error of the mean, minimum total length, maximum total length, and mean total length are shown along with the sample size, and the standard deviation from the mean of the relative weights for smallmouth bass in Pueblo Reservoir.*

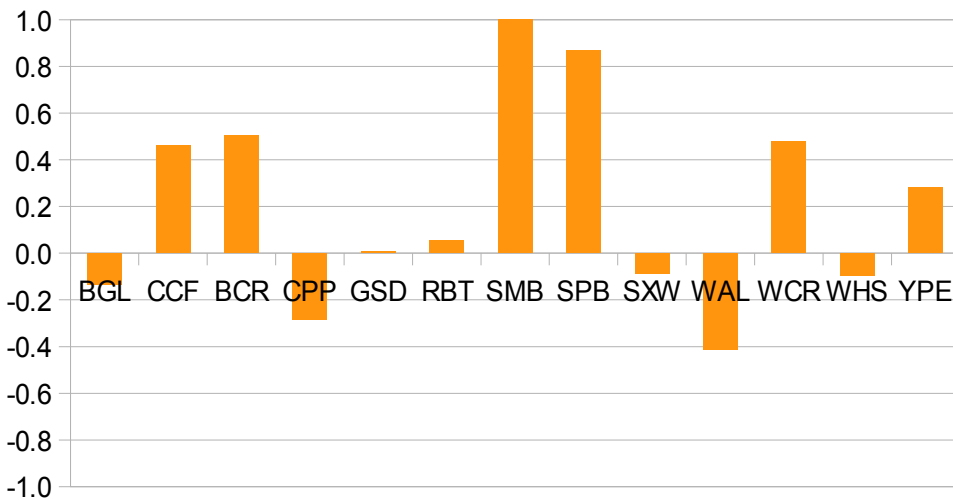

*Figure 7b: Correlation coefficients for annual mean relative weights between smallmouth bass (all length classes) and other species in Pueblo Reservoir from 1990 to 2011. Note positive correlations with species with significant overlaps in forage habits.*

Figure 7b shows the correlation of smallmouth bass Wr (all length classes combined) with other species in Pueblo Reservoir from 1990 to 2011. The correlations suggest substantial forage overlap with spotted bass, as well as some forage overlap with channel catfish, black crappie, white crappie, and yellow perch. Correlations with walleye, bluegill, and common carp are slightly negative. Figure 7c shows the Wr correlations between length classes of smallmouth bass with other species. Bluegill Wr are positively correlated with the two smaller length classes of smallmouth bass (below stock length and stock length), but negatively correlated with the quality and preferred combined length classes, suggesting that these length classes demonstrate significant predation on bluegill. Wr of all smallmouth bass length classes are positively correlated with Wr of channel catfish, suggesting substantial forage overlap. As expected the preferred and memorable combined length class is more strongly correlated with channel catfish, suggesting stronger commonality of prey. There are significant correlations between all three length groups of smallmouth bass with Wr of black crappie, with the smallest (below stock length) length class having the strongest correlation with black crappie, suggesting that smallmouth bass from 150 to 180 mm TL are less piscivorous than longer smallmouth bass. The stock and quality/preferred length groups have almost no correlation with Wr of common carp. The sub stock length group has a correlation of -0.496 with common carp, but this is hard to interpret because a predatory relationship is doubtful for bass in this length range.

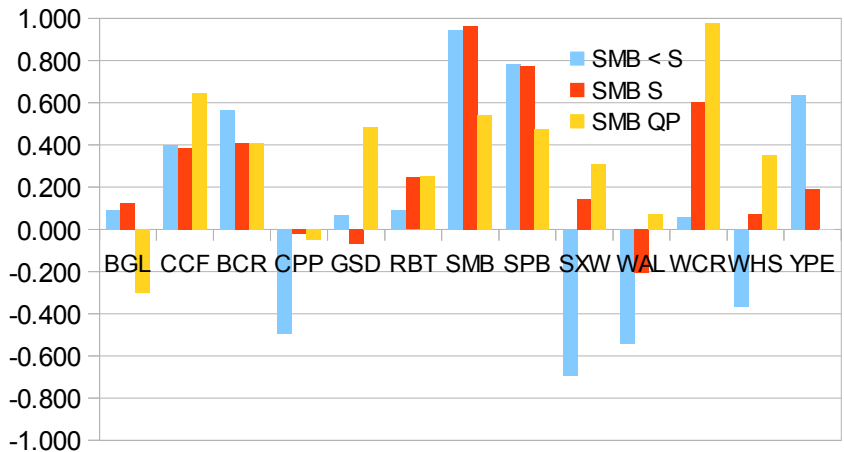

*Figure 7c: Correlation coefficients for annual mean relative weights between smallmouth bass (by length class) and other species in Pueblo Reservoir from 1990 to 2011. Note positive correlations with species with significant overlaps in forage habits.*

Only the Wr of the longest length group (quality/preferred) of smallmouth bass has a significant correlation (r = 0.488) with gizzard shad. Combined with the observation that this length group of smallmouth bass tends toward piscivory, this suggests support for the reproductive hypothesis that piscivorous fish will have Wr positively correlated with adult common carp, because adult common carp with high Wr have higher reproductive success thus produce more forage for piscivores. The small positive correlation of Wr with rainbow trout may represent some overlap of invertebrate food sources. All length groups of smallmouth bass have significant positive correlation with spotted bass. The smallest length group of smallmouth bass (TL 150 mm to 180 mm) has a significant negative overlap with the Wr of hybrid striped bass. Could hybrid striped bass actually be preying on smallmouth bass in this length range? It could also be that this length range of smallmouth bass have the same forage base as smaller bass which are prey for hybrid striped bass.

Negative correlations in Wr between walleye and smallmouth bass is consistent with the results of Wuellner et al. (2011) in four of six South Dakota lakes that smallmouth bass are not strongly competing with walleye for available forage. There is a large correlation (r = 0.978) between the annual Wr of white crappie and the longest group of smallmouth bass (TL > 280 mm). This should not be surprising since in most years, the sample includes a number of white crappie with TL > 300 mm, and correlations of white crappie Wr with other species have suggested significant piscivory. There is a significant correlation between the Wr of sub stock length smallmouth bass and yellow perch, suggesting forage overlap.

*Spotted Bass*

Annual relative weights (Wr) and associated metrics for spotted bass in Pueblo Reservoir are shown in Table 8a and Figure 8a. The mean relative weight over the time period is 108.0. In plump years, the mean Wr has been as high as 115.6; whereas, in lean years, the mean Wr has only dipped to 98.8.

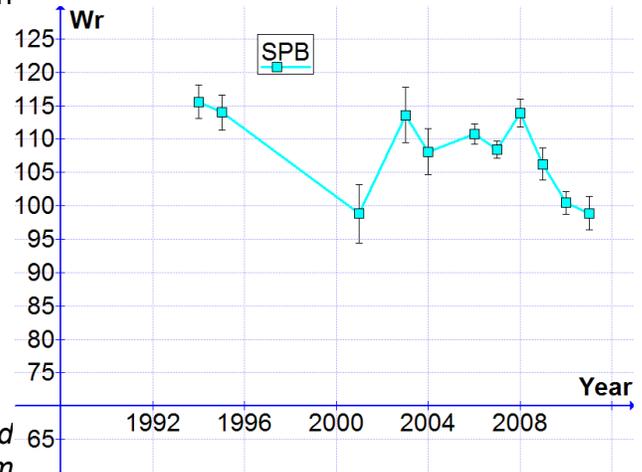

| SPB | Wr | SEM | TL min | TL max | TL mean | n | SD |
|---|---|---|---|---|---|---|---|
| 1994 | 115.6 | 2.5 | 150 | 380 | 214.8 | 32 | 13.87 |
| 1995 | 114.0 | 2.6 | 100 | 350 | 200.2 | 42 | 16.8 |
| 2001 | 98.8 | 4.4 | 245 | 370 | 321.3 | 4 | 8.765 |
| 2003 | 113.6 | 4.1 | 100 | 370 | 230.5 | 64 | 32.98 |
| 2004 | 108.1 | 3.5 | 185 | 380 | 300.8 | 6 | 8.486 |
| 2006 | 110.7 | 1.5 | 100 | 375 | 157.7 | 144 | 17.95 |
| 2007 | 108.4 | 1.3 | 110 | 395 | 202.8 | 43 | 15.52 |
| 2008 | 113.9 | 2.1 | 100 | 400 | 186.5 | 148 | 25.79 |
| 2009 | 106.2 | 2.4 | 190 | 395 | 290.7 | 23 | 11.35 |
| 2010 | 100.4 | 1.6 | 194 | 381 | 292.1 | 28 | 8.703 |
| 2011 | 98.9 | 2.5 | 195 | 374 | 256.8 | 8 | 7.093 |

*Table 8a/Figure 8a: Annual relative weights, standard error of the mean, minimum total length (mm), maximum total length (mm), and mean total length(mm) are shown along with the sample size, and the standard deviation from the mean of the relative weights for spotted bass in Pueblo Reservoir.*

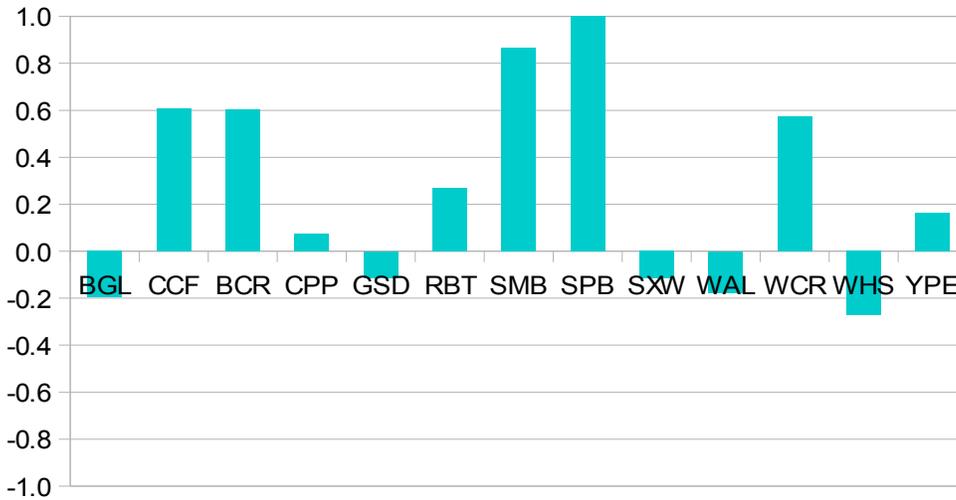

*Figure 8b: Correlation coefficients for annual mean relative weights between spotted bass (all length classes) and other species in Pueblo Reservoir from 1990 to 2011. Note positive correlations with species with significant overlaps in forage habits.*

Figure 8b shows the correlation of spotted bass Wr with other species in Pueblo Reservoir from 1994 to 2011. The correlations suggest substantial forage overlap with smallmouth bass ($p < 0.001$), as well as some forage overlap with channel catfish ($p = 0.024$), black crappie ($p = 0.140$), and white crappie ($p = 0.069$). The high correlation with smallmouth bass ($r = 0.867$) suggests a substantial forage overlap with this species; the fact that the mean relative weight is much higher in spotted bass than in smallmouth bass suggests that the spotted bass are getting more than their fair share of the forage and outcompeting smallmouth bass for available forage. The lack of any significant negative correlations suggests that spotted bass may be feeding primarily on something other than fish, probably crawfish and other benthic invertebrates.

*Hybrid Striped Bass*

Annual relative weights (Wr) and associated metrics for hybrid striped bass in Pueblo Reservoir are shown in Table 9a and Figure 9a. The mean relative weight over the time period is 91.4. In plump years, the mean Wr has been as high as 97.0; whereas, in lean years, the mean Wr has been as low as 85.3.

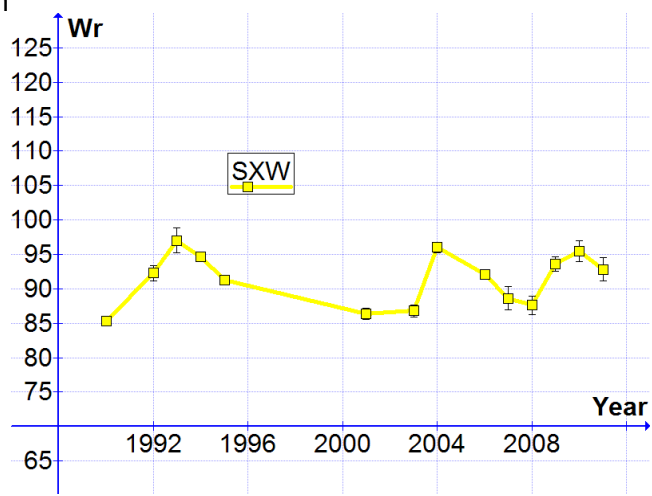

| SXW | Wr | SEM | TL min | TL max | TL mean | n | SD |
|---|---|---|---|---|---|---|---|
| 1990 | 85.3 | 0.5 | 53 | 520 | 321.0 | 156 | 6.5 |
| 1992 | 92.3 | 1.1 | 65 | 525 | 387.0 | 69 | 9.4 |
| 1993 | 97.0 | 1.8 | 275 | 552 | 448.3 | 85 | 16.3 |
| 1994 | 94.6 | 0.6 | 115 | 570 | 290.1 | 358 | 10.5 |
| 1995 | 91.3 | 0.4 | 85 | 630 | 340.0 | 146 | 9.3 |
| 2001 | 86.4 | 0.8 | 150 | 690 | 410.1 | 109 | 8.2 |
| 2003 | 86.8 | 0.9 | 325 | 690 | 456.9 | 59 | 6.8 |
| 2004 | 96.0 | 0.8 | 225 | 620 | 437.1 | 60 | 6.4 |
| 2006 | 92.1 | 0.7 | 140 | 615 | 416.0 | 78 | 6.2 |
| 2007 | 88.6 | 1.7 | 265 | 665 | 431.0 | 52 | 12.1 |
| 2008 | 87.6 | 1.3 | 410 | 615 | 522.1 | 24 | 6.2 |
| 2009 | 93.6 | 1.1 | 300 | 730 | 520.5 | 37 | 6.9 |
| 2010 | 95.5 | 1.5 | 197 | 650 | 415.8 | 25 | 7.6 |
| 2011 | 92.8 | 1.7 | 342 | 642 | 466.1 | 36 | 10.2 |

*Table 9a/Figure 9a: Annual relative weights, standard error of the mean, minimum total length (mm), maximum total length (mm), and mean total length(mm) are shown along with the sample size, and the standard deviation from the mean of the relative weights for hybrid striped bass in Pueblo Reservoir.*

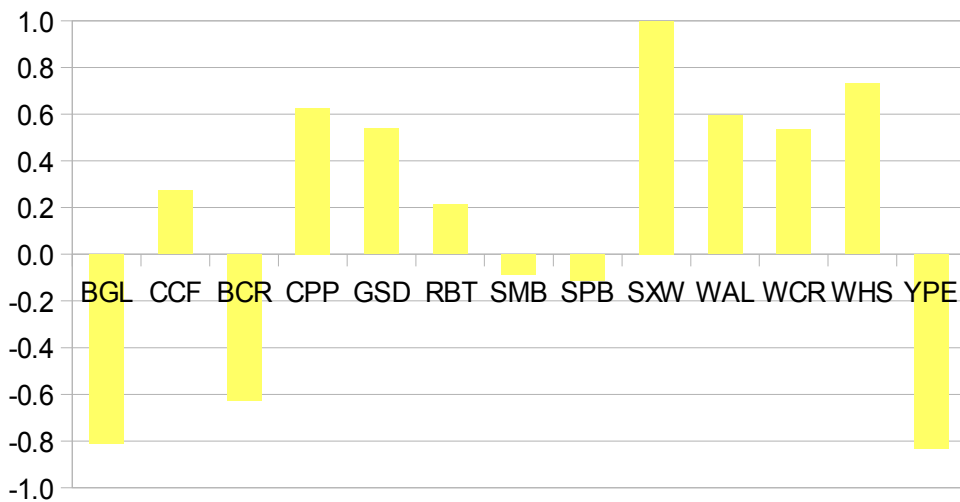

*Figure 9b: Correlation coefficients for annual mean relative weights between hybrid striped bass (all length classes) and other species in Pueblo Reservoir from 1990 to 2011. Note positive correlations with species with significant overlaps in forage habits.*

Figure 9b shows the correlation of hybrid striped bass Wr with other species in Pueblo Reservoir from 1990 to 2011. The most significant negative correlations are with important prey species: bluegill ($p = 004$), yellow perch ($p = 0.020$), and black crappie ($p = 0.128$). The most significant positive interactions are probably due to the fecundity of larger specimens of important prey species like gizzard shad ($p = 0.024$), white sucker ($p = 0.002$), and common carp ($p = 0.008$) being very high and hybrid striper bass having plenty to eat in years where these important prey species produce abundant offspring. The significant positive correlations that may be related to competition are with white crappie ($p = 0.070$) and with channel catfish ($p = 0.271$). These interactions can be explored in more detail by considering the correlations with different length classes of each species with hybrid striped bass.

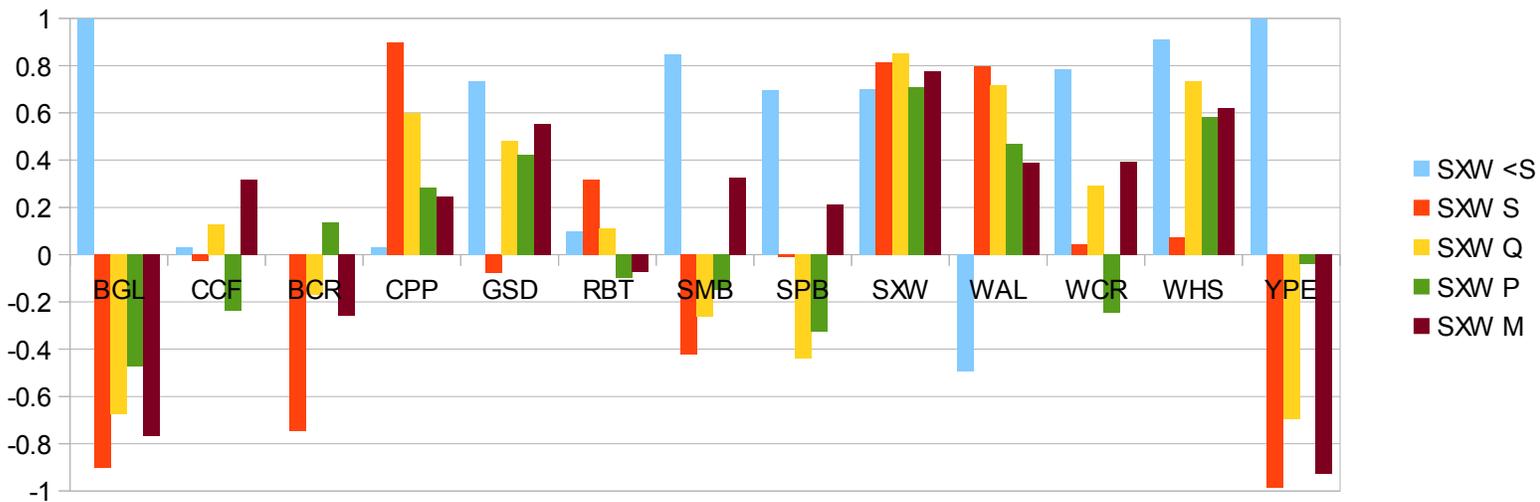

*Figure 9c: Correlation coefficients for annual mean relative weights between hybrid striped bass (by length class) and other species in Pueblo Reservoir from 1990 to 2011. Note positive correlations with species with significant overlaps in forage habits.*

Correlations of annual relative weights of different species are shown with different length classes of hybrid striped bass in Figure 9c. Since there were only four years with good numbers of hybrid white bass under 200 mm sampled, most of the correlations with sub stock length hybrid striped bass (SXW <S) are not significant. However, the large correlation with white sucker (p = 0.910) is significant (r = 0.045), suggesting either a significant overlap in forage, or (more likely) that the smaller hybrid striped bass feed heavily upon age zero white sucker (the fecundity hypothesis). The correlation between Wr of sub stock length hybrid striped bass with smallmouth bass is less significant (p = 0.075), but it is suggestive of likely competition.

Annual relative weight of stock length hybrid striped bass (SXW S, 200-300mm in length) has a number of significant correlations with other species, both positive and negative. The negative correlation with bluegill is both strong (r = -0.901) and significant (p = 0.018) suggesting that bluegill are most likely an important food source for this stock length hybrid striped bass. Likewise, the negative correlation with yellow perch is strong (r = -0.985) and significant (p = 0.015). The strong negative correlation with black crappie might indicate the importance of predation, but the significance is low (p = 0.231). The strongest positive correlation is with common carp (r = 0.898). This correlation is highly significant (p = 0.003), but it is not clear if this correlation suggests a heavy overlap of forage sources (sucking down the same benthos) or if stock length hybrids depend heavily on age zero carp as an important food source (the fecundity hypothesis). With a similar mix of pelagic and benthic food sources, it is no surprise that stock length hybrid striped bass have a strong (p = 0.796) and significant (r = 0.016) correlation with walleye.

As hybrid striped bass get longer up to the quality and preferred length classes, their forage dependence on bluegill, black crappie, age zero common carp, and yellow perch seem to decrease.

*White Crappie*

Annual relative weights (Wr) and associated metrics for white crappie in Pueblo Reservoir are shown in Table 10 and Figure 10a. The mean relative weight over the time period is 89.5, and the mean annual relative weights were below 90 most years, suggesting that white crappie in the reservoir tend to be on the thin side.

| WCR | Wr | SEM | TL min | TL max | TL mean | n | SD |
|---|---|---|---|---|---|---|---|
| 1992 | 89.9 | 3.0 | 135 | 280 | 178.9 | 18 | 12.6 |
| 1994 | 99.8 | 3.6 | 125 | 370 | 256.7 | 12 | 12.4 |
| 1995 | 100.7 | 5.5 | 130 | 300 | 206.5 | 13 | 19.8 |
| 2001 | 83.0 | 2.5 | 180 | 340 | 276.9 | 8 | 6.9 |
| 2003 | 84.4 | 3.4 | 140 | 365 | 238.0 | 5 | 7.7 |
| 2007 | 86.1 | 7.2 | 245 | 310 | 268.3 | 3 | 12.4 |
| 2008 | 85.9 | 1.6 | 235 | 305 | 271.3 | 12 | 5.7 |
| 2009 | 90.4 | 4.1 | 300 | 345 | 315.0 | 4 | 8.2 |
| 2010 | 85.4 | 3.0 | 317 | 329 | 323.0 | 2 | 4.3 |

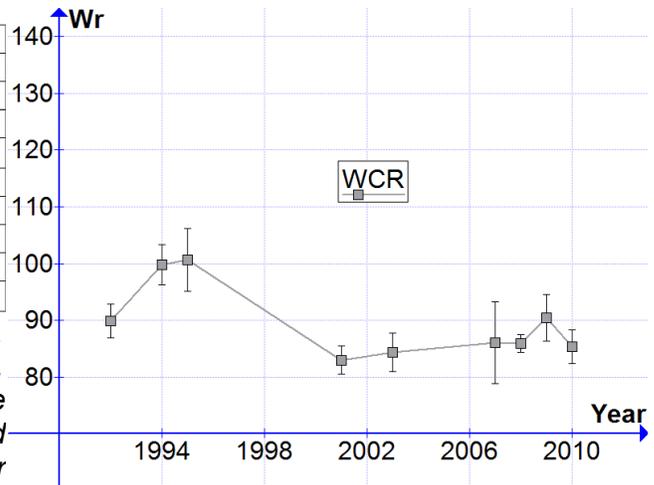

*Table 10/Figure 10a: Year to year relative weights, standard error of the mean, minimum total length, maximum total length, and mean total length are shown along with the sample size, and the standard deviation from the mean of the relative weights for white crappie in Pueblo Reservoir.*

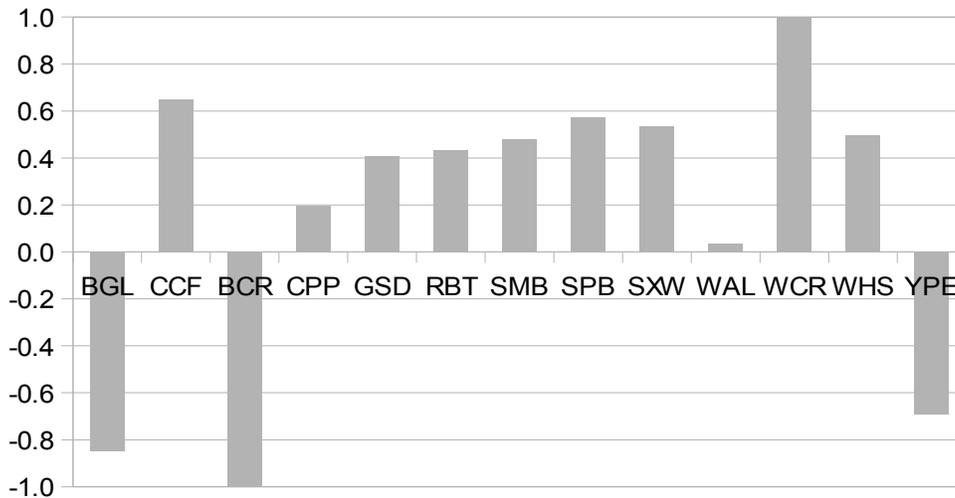

*Figure 10b: Correlation coefficients for annual mean relative weights between white crappie and other species in Pueblo Reservoir from 1992 to 2010. Note positive correlations with species with significant overlaps in forage habits (channel catfish, smallmouth bass, spotted bass, and hybrid striped bass) and significant negative correlations with prey species (bluegill and yellow perch).*

The strong (r = -0.848) and significant (p = 0.016) negative correlation of relative weights with bluegill suggest that the interspecies relationship with bluegill is more one of predation than competition for food. Ellison (1984) showed that white crappie begin to be efficient piscivores at 150 mm TL, which can be reached by age 2, and that by 200 mm TL (age 3 or 4), fish make up over 50% of their diet. Elliston (1984) expected white crappie greater than 200 mm TL would be an efficient predator on young of the year bluegill in clear waters, and this seems to be true here. The negative correlation of relative weight between white crappie and yellow perch is not as strong (r = -0.695) nor as significant (p = 0.063), but it still suggests more of a predatory than a competitive relationship between the piscivorous white crappie and yellow perch. The correlation between relative weights of white crappie and black crappie is not significant (p = 1.000) because there are only two years when relative weight data are available for both.

*Black Crappie*

Annual relative weights (Wr) and associated metrics for black crappie in Pueblo Reservoir are shown in Table 11 and Figure 11a. Black crappie are a recent introduction to the reservoir, first appearing in survey data in 2004. The mean relative weight over the years is 105.4, suggesting black crappie are able to find a niche and eat well at relatively modest stocking densities.

| BCR | Wr | SEM | TL min | TL max | TL mean | n | SD |
|---|---|---|---|---|---|---|---|
| 2004 | 102.8 | 22.5 | 165 | 310 | 220.0 | 5 | 50.2 |
| 2006 | 92.6 | 6.9 | 120 | 225 | 178.8 | 4 | 13.7 |
| 2008 | 130.7 | 10.9 | 100 | 275 | 132.5 | 14 | 40.8 |
| 2011 | 95.5 | 8.0 | 216 | 251 | 235.0 | 3 | 13.8 |

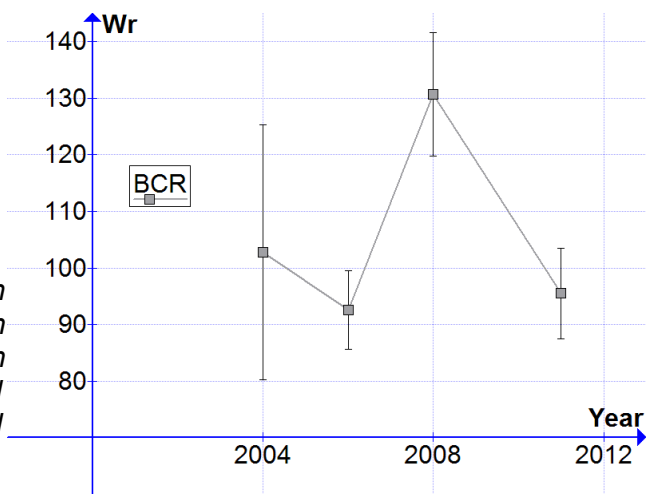

Table 11/Figure 11a: Year to year relative weights, standard error of the mean, minimum total length, maximum total length, and mean total length are shown along with the sample size, and the standard deviation from the mean of the relative weights for black crappie in Pueblo Reservoir. Due to small sample sizes and variation in relative weight among samples, the standard errors in the mean tend to be large.

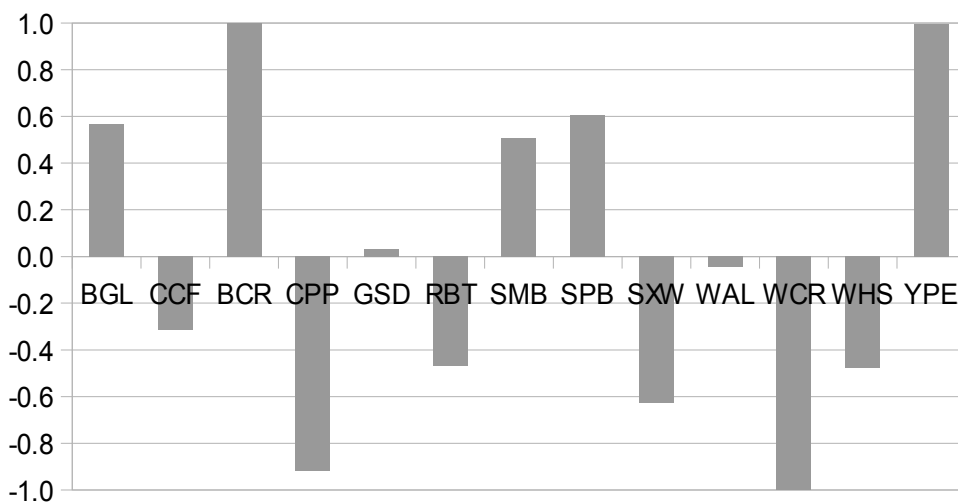

Figure 11b: Correlation coefficients for annual mean relative weights between black crappie and other species in Pueblo Reservoir from 2004 to 2011. Note positive correlations with species with significant overlaps in forage habits (bluegill, smallmouth bass, spotted bass, and yellow perch).

As might be expected, Wr of black crappie have positive correlations with species generally considered strong food competitors: bluegill, smallmouth bass, spotted bass, and yellow perch. However, due to the small number of years with available data, these observations only have a good level of significance (p = 0.032) for yellow perch. Significance levels for bluegill (p = 0.217), smallmouth bass (p = 0.192), and spotted bass (p = 0.140) are marginal to poor.

    The negative correlation with common carp (r = -0.919) is both large and significant (p = 0.014). However, this is a challenge to understand. If black crappie were feeding strongly on age zero common carp, the correlation should be positive (the fecundity hypothesis) as it is for other species suggested to be feeding strongly on age zero common carp. It is likely that the fecundity hypothesis contributes to the strong negative correlation through competition. High Wr in adult common carp contributes to a strong age zero class of common carp which competes strongly with black crappie for available zooplankton.

*Gizzard Shad*

Annual relative weights (Wr) and associated metrics for gizzard shad in Pueblo Reservoir are shown in Table 12 and Figure 12a. The mean annual relative weight of gizzard shad was 88.9, suggesting that the gizzard shad tend to be well populated in the reservoir relative to their food sources. The abundance of adult gizzard shad may be due to a lack of natural predators once the shad are over 300 mm in total length. Willis (1987) found that the mean relative weight of adult gizzard shad was strongly correlated with the number of available age zero gizzard shad.

| GSD | Wr | SEM | TL min | TL max | TL mean | n | SD |
|---|---|---|---|---|---|---|---|
| 1990 | 85.7 | 0.7 | 290 | 405 | 346.8 | 112 | 7.4 |
| 1992 | 86.4 | 0.7 | 205 | 405 | 346.8 | 133 | 7.7 |
| 1993 | 101.3 | 1.4 | 246 | 473 | 342.9 | 90 | 19.1 |
| 1994 | 85.8 | 1.2 | 225 | 395 | 343.8 | 214 | 17.4 |
| 1995 | 94.2 | 1.3 | 180 | 395 | 305.4 | 185 | 17.3 |
| 2001 | 90.7 | 1.2 | 270 | 410 | 344.9 | 71 | 10.3 |
| 2003 | 81.0 | 0.7 | 180 | 415 | 353.7 | 191 | 9.0 |
| 2004 | 106.4 | 1.8 | 180 | 410 | 280.4 | 80 | 15.9 |
| 2006 | 80.4 | 0.6 | 180 | 425 | 323.9 | 154 | 7.9 |
| 2007 | 85.2 | 1.3 | 195 | 420 | 297.5 | 80 | 11.3 |
| 2008 | 84.8 | 0.7 | 195 | 420 | 322.3 | 261 | 10.9 |
| 2009 | 90.8 | 1.6 | 260 | 425 | 344.3 | 28 | 8.5 |
| 2010 | 88.9 | 2.0 | 265 | 444 | 355.8 | 103 | 20.1 |
| 2011 | 83.4 | 1.8 | 330 | 475 | 375.0 | 50 | 12.9 |

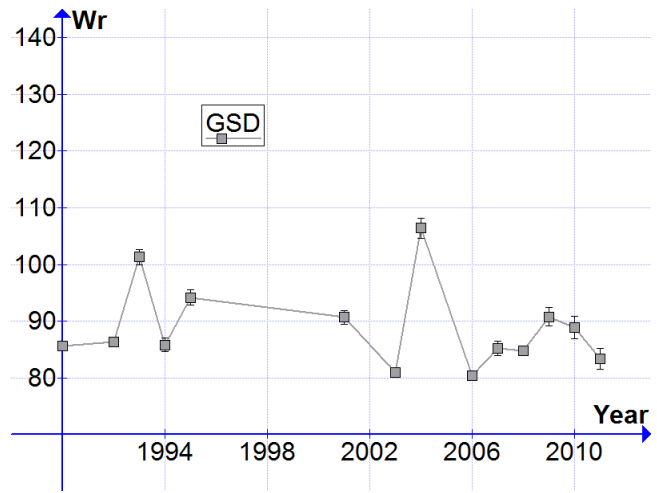

*Table 12/Figure 12a: Year to year relative weights, standard error of the mean, minimum total length, maximum total length, and mean total length are shown along with the sample size, and the standard deviation from the mean of the relative weights for gizzard shad in Pueblo Reservoir.*

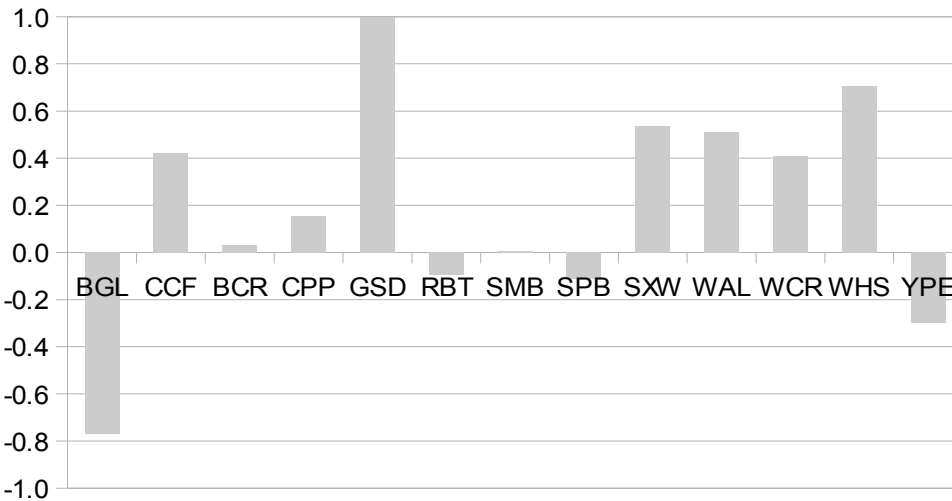

*Figure 12b: Correlation coefficients for annual mean relative weights between gizzard shad and other species in Pueblo Reservoir from 1990 to 2011. Note positive correlations with species with significant overlaps in forage habits (white sucker). Correlations with some predators may be positive due to the fecundity hypothesis.*

Bluegill are well known to be strong food competitors with gizzard shad (Aday et al., 2003). The correlation is strongly negative in this case because the bluegill are not strongly competing with the length classes of gizzard shad sampled in the survey (TL > 180 mm), but rather with their age zero offspring. High mean Wr in adult gizzard shad leads to abundant age zero cohorts. These age zero cohorts compete strongly with bluegill and reduce the mean bluegill Wr. The weaker negative correlation with relative weights of yellow perch seems to agree with the findings of Roseman et al. (1996) that age zero gizzard shad do not compete as strongly with yellow perch as with bluegill.

*Common Carp*

Annual relative weights (Wr) and associated metrics for common carp in Pueblo Reservoir are shown in Table 13 and Figure 13a. The mean annual relative weight of common carp was 86.8, suggesting that the common carp tend to be well populated in the reservoir relative to their food sources. The abundance of adult common carp may be due to a lack of natural predators once they are over 300 mm in total length. Since adults of the species are plentiful relative to available food sources, fecundity may be limited by body condition with plentiful offspring in years with high relative weights.

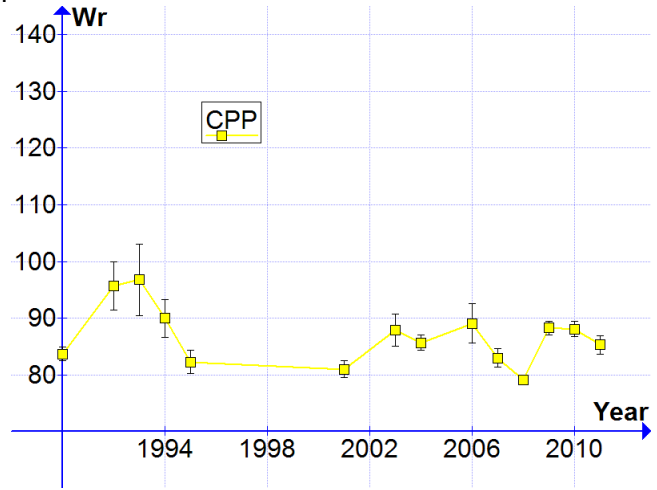

| CPP | Wr | SEM | TL min | TL max | TL mean | n | SD |
|---|---|---|---|---|---|---|---|
| 1990 | 83.7 | 1.2 | 280 | 630 | 543.7 | 23 | 5.7 |
| 1992 | 95.7 | 4.3 | 570 | 640 | 598.3 | 6 | 10.5 |
| 1993 | 96.8 | 6.3 | 455 | 622 | 505.5 | 10 | 20.0 |
| 1994 | 90.0 | 3.3 | 210 | 715 | 498.1 | 29 | 18.0 |
| 1995 | 82.3 | 2.1 | 360 | 710 | 581.6 | 31 | 11.5 |
| 2001 | 81.0 | 1.5 | 430 | 670 | 569.5 | 10 | 4.7 |
| 2003 | 87.9 | 2.8 | 500 | 620 | 556.8 | 25 | 13.8 |
| 2004 | 85.7 | 1.4 | 510 | 730 | 580.6 | 16 | 5.5 |
| 2006 | 89.1 | 3.5 | 285 | 630 | 498.8 | 25 | 17.4 |
| 2007 | 83.0 | 1.6 | 280 | 670 | 527.9 | 21 | 7.4 |
| 2008 | 79.1 | 0.8 | 270 | 755 | 526.9 | 40 | 5.2 |
| 2009 | 88.3 | 1.2 | 455 | 635 | 531.8 | 11 | 4.0 |
| 2010 | 88.1 | 1.3 | 479 | 667 | 570.9 | 14 | 4.7 |
| 2011 | 85.3 | 1.6 | 439 | 662 | 571.2 | 13 | 6.1 |

*Table 13/Figure 13a: Year to year relative weights, standard error of the mean, minimum total length, maximum total length, and mean total length are shown along with the sample size, and the standard deviation from the mean of the relative weights for common carp in Pueblo Reservoir.*

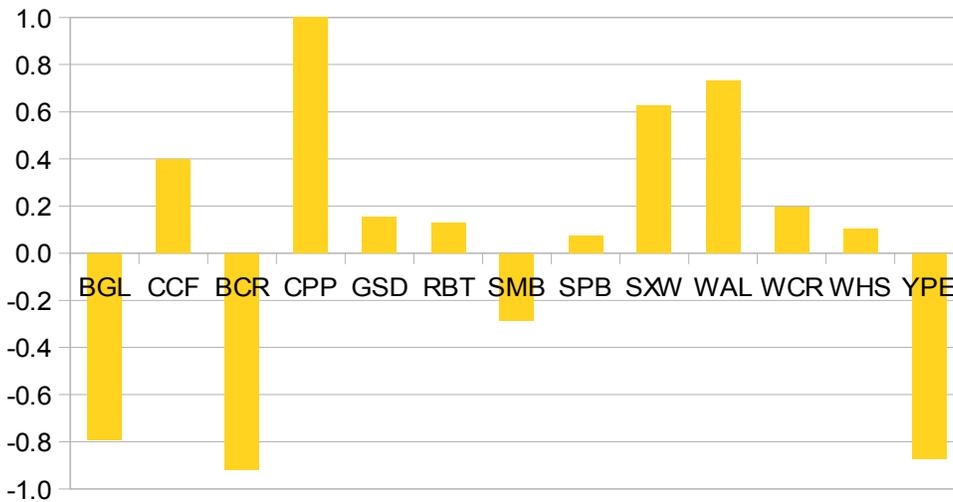

*Figure 13b: Correlation coefficients for annual mean relative weights between common carp and other species in Pueblo Reservoir from 1990 to 2011. Since sampled common carp are adults, most correlations (positive and negative) may have contributions from the fecundity hypothesis. Age zero common carp are probably strong food competitors with bluegill, black crappie, and yellow perch.*

Interpreting Fig. 5h with the fecundity hypothesis suggests that stock and quality length walleye make strongest use of age zero carp with the preferred and memorable length classes of walleye making more use of age zero gizzard shad. Interpreting Fig. 7c with the fecundity hypothesis suggests that smallmouth bass below stock length are in competition with age zero carp for food and that the quality and preferred length classes of smallmouth bass feed heavily on age zero gizzard shad but do not make strong use of age zero carp. Interpreting Fig 9c with the fecundity hypothesis suggests that stock length hybrid striped bass feed more strongly than longer length classes on age zero carp, and that the preferred and memorable length classes of hybrid striped bass make more use of age zero gizzard shad.

*White Sucker*

Annual relative weights (Wr) and associated metrics for white sucker in Pueblo Reservoir are shown in Table 14 and Figure 14a. The mean annual relative weight of white sucker was 97 suggesting white sucker have abundant food most years relative to their population.

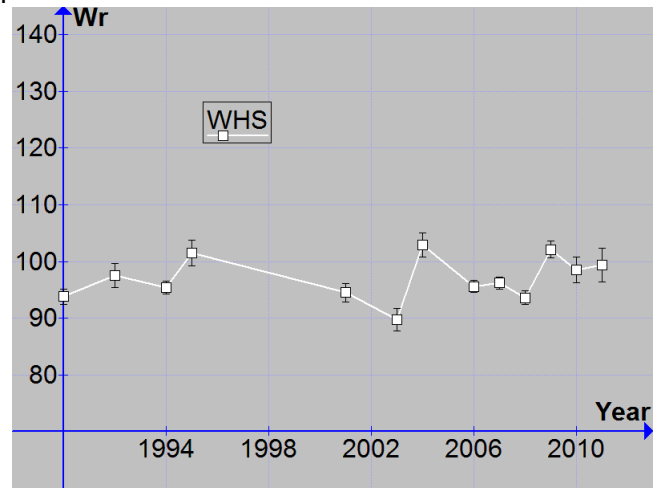

| WHS | Wr | SEM | TL min | TL max | TL mean | n | SD |
|---|---|---|---|---|---|---|---|
| 1990 | 93.8 | 1.4 | 330 | 520 | 418.7 | 45 | 9.1 |
| 1992 | 97.5 | 2.1 | 320 | 500 | 422.3 | 30 | 11.4 |
| 1994 | 95.4 | 1.1 | 160 | 480 | 392.6 | 69 | 9.5 |
| 1995 | 101.5 | 2.3 | 215 | 480 | 393.8 | 20 | 10.5 |
| 2001 | 94.5 | 1.6 | 325 | 475 | 382.1 | 42 | 10.1 |
| 2003 | 89.8 | 2.0 | 195 | 470 | 367.6 | 21 | 9.1 |
| 2004 | 102.9 | 2.1 | 175 | 485 | 386.7 | 24 | 10.3 |
| 2006 | 95.6 | 1.1 | 200 | 490 | 371.1 | 46 | 7.8 |
| 2007 | 96.2 | 1.1 | 180 | 490 | 400.4 | 41 | 6.8 |
| 2008 | 93.6 | 1.2 | 170 | 485 | 382.6 | 44 | 8.1 |
| 2009 | 102.1 | 1.5 | 340 | 485 | 421.7 | 21 | 7.1 |
| 2010 | 98.5 | 2.3 | 384 | 496 | 436.1 | 15 | 8.7 |
| 2011 | 99.4 | 3.0 | 354 | 497 | 453.0 | 31 | 16.8 |

*Table 14/Figure 14a: Year to year mean relative weights, standard error of the mean, minimum total length, maximum total length, and mean total length are shown along with the sample size, and the standard deviation from the mean of the relative weights for white sucker in Pueblo Reservoir.*

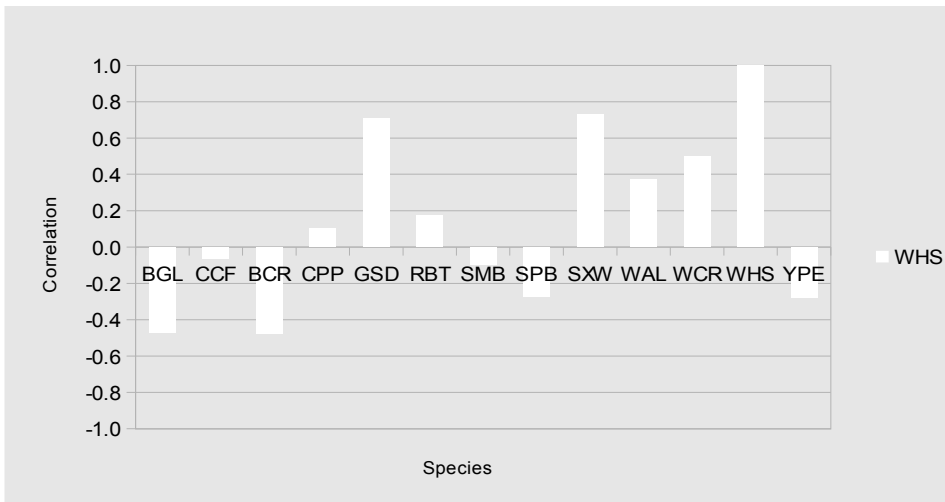

*Figure 14b: Correlation coefficients for annual mean relative weights between white sucker and other species. Since sampled white sucker are adults, some correlations (positive and negative) may have contributions from the fecundity hypothesis. The large correlation of mean annual Wr between white sucker and gizzard shad suggests commonality of forage sources. Perhaps gizzard shad in Pueblo Reservoir are making more use of benthic food sources, even though they are a pelagic feeder in most ecosystems.*

Interpreting Fig. 5h with the fecundity hypothesis suggests that while walleye of all length classes may make some use of age zero white sucker, the preferred and memorable length classes seem to be making the strongest use of white sucker as forage. Interpreting Fig. 7c with the fecundity hypothesis suggests that smallmouth bass below stock length are food competitors with age zero white sucker, and that the quality and preferred length classes may be preying upon age zero white sucker. However, Tables A3 and A4 show that neither the correlations nor the p-values are compelling. Interpreting Fig. 9c with the fecundity hypothesis shows that hybrid striped bass below stock length are likely preying strongly upon age zero white sucker, and that quality, preferred, and memorable length classes of hybrid striped bass are also making use of age zero white sucker as an important forage source. The correlations of mean annual Wr between white sucker and these length classes of hybrid striped bass are all significant at the level of $p < 0.05$.

*Rainbow Trout*

Annual relative weights (Wr) and associated metrics for rainbow trout in Pueblo Reservoir are shown in Table 15 and Figure 15a. The mean annual relative weight of rainbow trout was 84.8 suggesting that rainbow trout are not competing well for food in the reservoir. The reservoir is a bit on the warm side and also does not support the primary production levels best suitable for rainbow trout. One wonders if the CDPW stocks rainbow trout due to angler expectations for Colorado waters rather than a good match of the species to the characteristics of the reservoir.

| RBT  | Wr   | SEM | TL min | TL max | TL mean | n | SD   |
|------|------|-----|--------|--------|---------|---|------|
| 1994 | 92.2 | 6.8 | 325    | 455    | 387.5   | 4 | 13.6 |
| 2003 | 73.7 | 2.3 | 280    | 310    | 293.3   | 3 | 3.9  |
| 2004 | 87.7 | 5.1 | 260    | 425    | 318.9   | 9 | 15.2 |
| 2006 | 92.7 | 5.4 | 350    | 490    | 390.0   | 4 | 10.9 |
| 2009 | 80.0 | 1.3 | 265    | 270    | 267.5   | 2 | 2.5  |
| 2010 | 82.8 | 0.7 | 207    | 396    | 273.3   | 3 | 1.2  |

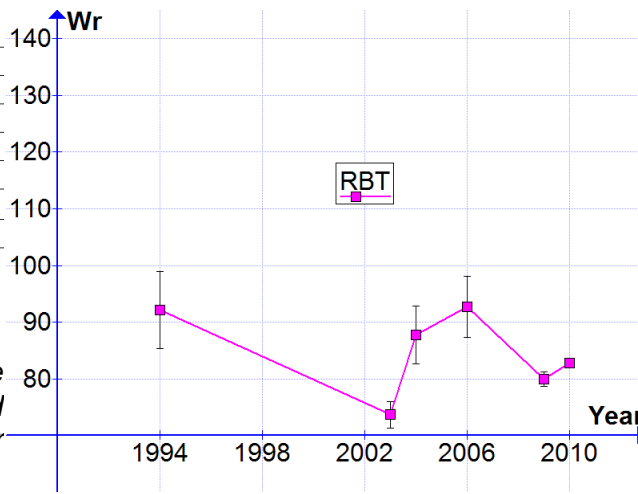

*Table 15/Figure 15a: Year to year relative weights, standard error of the mean, minimum total length, maximum total length, and mean total length are shown along with the sample size, and the standard deviation from the mean of the relative weights for rainbow trout in Pueblo Reservoir.*

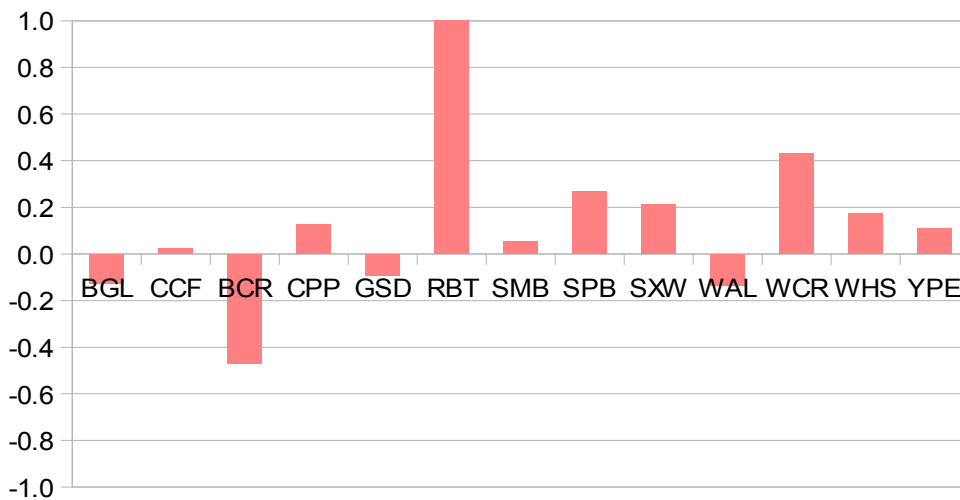

*Figure 15b: Correlation coefficients for annual mean relative weights between white sucker and other species in Pueblo Reservoir from 1994 to 2010. None of the correlations are significant above the p = 0.01 level. This is not surprising since the numbers of rainbow trout in the reservoir are probably too low to have significant competitive or predatory relationships with other species.*

**Discussion of Findings Regarding Pueblo Reservoir Fisheries**
The large biomass of rough fish in Pueblo reservoir probably prevents the fishery from being as productive as it could be for more desirable species. The energy requirements to maintain such a large biomass of adult common carp, white sucker, and gizzard shad limit growth and production of walleye and hybrid striped bass. Angler satisfaction adds considerable management challenges as the CDPW seems to be giving into the temptation to stock larger numbers of walleye and hybrid striped bass than can grow quickly under the available conditions. In future years, managers will face the additional management challenge of the proliferation of zebra mussels.

Food web dynamics at Pueblo Reservoir are complex. The study of correlations of mean annual relative weights has provided some additional insights. Most of the piscivorous species seem dependent on strong age zero classes of gizzard shad, common carp, and white sucker. The strength of these age zero classes seems to depend more strongly on the relative weights of these species (via the fecundity hypothesis) rather than the number of adults of these species. However, of walleye, hybrid striped bass, smallmouth bass, and channel catfish, all except for channel catfish demonstrate decreasing relative weight with length and all demonstrate mean relative weights far below 100 most years, with the average Wr over the years being below 90. The strongest mitigating factor of these thin, hungry piscivore populations as been the stocking of bluegill and black crappie which has provided a one year bump in their relative weights, but is not sustainable.

Clearly there are no easy management answers to the problem of adequate forage for the populations of hungry piscivores. In hindsight, originally populating the reservoir with threadfin shad rather than gizzard shad may have been a better choice to prevent such a large population of shad from growing beyond susceptibility to predation. Stocking northern pike and/or tiger muskellunge is being done in other reservoirs on an experimental basis in hopes of reducing populations of mature suckers. Stocking yellow perch to date has not been able to create a self-sustaining population sufficient for providing adequate forage to the desired walleye population and carries the additional risk of bucket biologists transplanting yellow perch to nearby coldwater reservoirs and significantly damaging excellent trout fisheries along the front range.

It seems that Pueblo Reservoir is already a strongly benthic/demersal system lacking sufficient primary and secondary pelagic production and benthic to pelagic coupling to maintain the desired growth rates at current stock densities of walleye and hybrid striped bass. Zebra mussel infestation in other waters is well known to couple primary pelagic production into the benthos and present further challenges to pelagic production. At the same time, even species such as catfish and freshwater drum that are capable of directly consuming zebra mussels have not seen significant increases in populations or body condition following zebra mussel infestation, presumably because consuming zebra mussels is seldom energetically attractive compared with making use of other available forage. Therefore, we expect that the anticipated infestation of zebra mussels will negatively impact production of gizzard shad and the piscivores which depend on them, thus reducing production of walleye, hybrid striped bass, spotted bass, smallmouth bass, and possibly also channel catfish. However, channel catfish may be enough of a generalist not to suffer greatly.

There are potential benefits in efforts to reduce the populations of adult common carp, gizzard shad, and white sucker. Educating anglers of the need to reduce these species is a first step, because anglers tend to instinctively believe they are benefiting the system by releasing fish they do not intend to eat, and there is a significant bycatch of common carp among walleye and bass anglers. There may also be an opportunity to interest ethnic populations of Pueblo in the harvest and consumption of species considered "rough fish" by other anglers. If recreational angling cannot sufficiently reduce the populations of adults of these species, one wonders if commercial harvest might be effectively combined with available local markets.

**Discussion of Method Using Relative Weight Correlations**
Porath and Peters (1997) had previously shown that relative weights in walleye can be a cost effective method to assess prey availability. This study demonstrates that correlations among annual mean relative weights can serve as a cost effective method for studying interspecies competition and predation on a larger scale. In the long term, correlations among mean annual relative weights are unlikely to replace the valuable information gained from studies of stomach contents and more traditional studies of stock densities and isotope analyses. However, if weight and length measurements are available for multiple species for a sufficient number of years, then additional information regarding the food web dynamics may be gleaned without the exhaustive and expensive stock sampling and stomach content analysis that has been the traditional approach to unraveling complex food web dynamics.

Budget and manpower constraints likely will not often allow the detailed study of food web dynamics analogous to comprehensive studies at Blue Mesa Reservoir, Colorado (Johnson and Koski, 2005) or Spirit Lake, Iowa (Liao et al., 2001; Liao et al., 2002). However, weight and length type surveys are commonly employed yearly or every other year to assess fishery populations, status, and needs on a much wider array of lakes and reservoirs than permit detailed food web studies. The present study demonstrates a method for re-assessing available data to suggest likely food web interactions in greater detail than other methods with minimal cost and labor requirements.

At the same time, it is essential that in the bureaucratic challenge to provide sound management with limited resources, scientists and managers not succumb to the temptation to become overly confident in conclusions ultimately demonstrating only correlation without compelling evidence for causality. For example, many plausible explanations offered in the results section above depend on the fecundity hypothesis, that is, the hypothesis that relative weights between adults in a prey species are positively and significantly correlated with relative weights in the predator species because the high body condition reflected in relative weight leads to a strong age zero class of the prey species, thus creating a higher body condition among predator species. To our knowledge, this mechanism has only been explicitly shown in gizzard shad in Kansas reservoirs (Willis, 1987) and is far from being established as a general mechanism relating relative weights of adults of a prey species with relative weights of their predators.

There is an undeniable appeal in the prospect of having simple indicators of the degree of competition and predation as might be suggested by the correlation coefficients. However, exploration of this possible identification is in its infancy and is fraught with a number of potential confounding factors. Even if there remains a general trend of likely association after confirmation in additional ecosystems and evidence based confirmation in Pueblo Reservoir, the authors of the present study do not think a correlation coefficient in isolation suggests more than a probability of an underlying competitive or predatory relationship and an estimate of the interaction strength between species.

Many mechanistic and relational details discussed in the above results section may sound quite reasonable, but require further empirical validation before being elevated to verified principles of general understanding. However tentatively such knowledge is held at the present time (or at the time a similar study is performed elsewhere), it may represent a valuable perspective for managers to base decisions until more certain information becomes available at a later time.

More work is needed to suggest a practical approach for combining correlation coefficients as estimates of interaction strength into models of population and food web dynamics. The present study may be a small step toward exploring "new ways to estimate biologically reasonable model coefficients from empirical data ..." (Berlow et al., 2004), or it might be an idea that seems clever at first but is precluded from fruitful applications by unmanageable implementation details. Even if correlation coefficients cannot be directly transformed into interaction matrix elements (or an analogous interaction strengths), they should provide a reality check for any independent methods purporting to quantitatively estimate interaction strengths among fishes. Furthermore, it may be that condition indexes (some measure of plumpness, energy intake, etc.) can be developed for taxa other than fishes so that the approach of the present study may be applicable to broader ecosystems. The present study also suggests both a need and the ability to consider the differing interaction strengths for different length/different life stages, as Pimm and Rice (1987) also have suggested. It may be noteworthy that if one takes the correlation coefficient as a measure of interaction strength, there appear to be more relatively strong interactions that might have been supposed from the suggestion of Berlow et al. (2004) that there are few strong interactions and many weak interactions in most ecosystems.

**Appendix**

|  | BGL | CCF | BCR | CPP | GSD | RBT | SMB | SPB | SXW | WAL | WCR | WHS | YPE |
|---|---|---|---|---|---|---|---|---|---|---|---|---|---|
| BGL | 1.000 | -0.697 | 0.567 | -0.791 | -0.770 | -0.126 | -0.133 | -0.196 | -0.812 | -0.886 | -0.848 | -0.468 | 0.742 |
| CCF | -0.697 | 1.000 | -0.315 | 0.399 | 0.420 | 0.025 | 0.461 | 0.608 | 0.271 | 0.464 | 0.648 | -0.066 | -0.732 |
| BCR | 0.567 | -0.315 | 1.000 | -0.919 | 0.029 | -0.470 | 0.506 | 0.604 | -0.628 | -0.045 | -1.000 | -0.477 | 0.995 |
| CPP | -0.791 | 0.399 | -0.919 | 1.000 | 0.155 | 0.126 | -0.286 | 0.075 | 0.626 | 0.731 | 0.197 | 0.103 | -0.872 |
| GSD | -0.770 | 0.420 | 0.029 | 0.155 | 1.000 | -0.094 | 0.006 | -0.113 | 0.537 | 0.512 | 0.407 | 0.705 | -0.302 |
| RBT | -0.126 | 0.025 | -0.470 | 0.126 | -0.094 | 1.000 | 0.055 | 0.268 | 0.212 | -0.136 | 0.431 | 0.173 | 0.112 |
| SMB | -0.133 | 0.461 | 0.506 | -0.286 | 0.006 | 0.055 | 1.000 | 0.867 | -0.088 | -0.415 | 0.480 | -0.098 | 0.279 |
| SPB | -0.196 | 0.608 | 0.604 | 0.075 | -0.113 | 0.268 | 0.867 | 1.000 | -0.115 | -0.175 | 0.572 | -0.272 | 0.165 |
| SXW | -0.812 | 0.271 | -0.628 | 0.626 | 0.537 | 0.212 | -0.088 | -0.115 | 1.000 | 0.595 | 0.533 | 0.730 | -0.834 |
| WAL | -0.886 | 0.464 | -0.045 | 0.731 | 0.512 | -0.136 | -0.415 | -0.175 | 0.209 | 1.000 | 0.036 | 0.372 | -0.492 |
| WCR | -0.848 | 0.648 | -1.000 | 0.197 | 0.407 | 0.431 | 0.480 | 0.572 | 0.533 | 0.036 | 1.000 | 0.497 | -0.695 |
| WHS | -0.468 | -0.066 | -0.477 | 0.103 | 0.705 | 0.173 | -0.098 | -0.272 | 0.730 | 0.372 | 0.497 | 1.000 | -0.280 |
| YPE | 0.742 | -0.732 | 0.995 | -0.872 | -0.302 | 0.112 | 0.279 | 0.165 | -0.834 | -0.492 | -0.695 | -0.280 | 1.000 |

*Table A1: Correlation coefficients between annual relative weights for thirteen species sampled from Pueblo Reservoir, Colorado from 1990 to 2011. Color codes represent range of p-values. Magenta represents p < 0.01; green, p < 0.05; yellow, p < 0.1; red, p < 0.2.*

|  | BGL | CCF | BCR | CPP | GSD | RBT | SMB | SPB | SXW | WAL | WCR | WHS | YPE |
|---|---|---|---|---|---|---|---|---|---|---|---|---|---|
| BGL |  | 0.018 | 0.217 | 0.006 | 0.008 | 0.394 | 0.377 | 0.321 | 0.004 | 0.001 | 0.016 | 0.121 | 0.046 |
| CCF | 0.018 |  | 0.303 | 0.079 | 0.067 | 0.471 | 0.056 | 0.024 | 0.174 | 0.047 | 0.030 | 0.415 | 0.049 |
| BCR | 0.217 | 0.303 |  | 0.014 | 0.482 | 0.344 | 0.192 | 0.140 | 0.128 | 0.471 | 1.000 | 0.208 | 0.032 |
| CPP | 0.006 | 0.079 | 0.014 |  | 0.298 | 0.356 | 0.172 | 0.404 | 0.008 | 0.001 | 0.306 | 0.369 | 0.012 |
| GSD | 0.008 | 0.067 | 0.482 | 0.298 |  | 0.392 | 0.492 | 0.370 | 0.024 | 0.031 | 0.138 | 0.004 | 0.280 |
| RBT | 0.394 | 0.471 | 0.344 | 0.356 | 0.392 |  | 0.440 | 0.243 | 0.266 | 0.345 | 0.123 | 0.305 | 0.429 |
| SMB | 0.377 | 0.056 | 0.192 | 0.172 | 0.492 | 0.440 |  | 0.000 | 0.387 | 0.079 | 0.095 | 0.375 | 0.296 |
| SPB | 0.321 | 0.024 | 0.140 | 0.404 | 0.370 | 0.243 | 0.000 |  | 0.368 | 0.303 | 0.069 | 0.209 | 0.377 |
| SXW | 0.004 | 0.174 | 0.128 | 0.008 | 0.024 | 0.266 | 0.387 | 0.368 |  | 0.237 | 0.070 | 0.002 | 0.020 |
| WAL | 0.001 | 0.047 | 0.471 | 0.001 | 0.031 | 0.345 | 0.079 | 0.303 | 0.237 |  | 0.463 | 0.105 | 0.044 |
| WCR | 0.016 | 0.030 | 1.000 | 0.306 | 0.138 | 0.123 | 0.095 | 0.069 | 0.070 | 0.463 |  | 0.087 | 0.063 |
| WHS | 0.121 | 0.415 | 0.208 | 0.369 | 0.004 | 0.305 | 0.375 | 0.209 | 0.002 | 0.105 | 0.087 |  | 0.295 |
| YPE | 0.046 | 0.049 | 0.032 | 0.012 | 0.280 | 0.429 | 0.296 | 0.377 | 0.020 | 0.044 | 0.063 | 0.295 |  |

*Table A2: p-values for correlation coefficients shown in Table A1.*

|  | SMB <S | SMB S | SMB QP | WAL <S | WAL S | WAL Q | WAL PM | SXW <S | SXW S | SXW Q | SXW P | SXW M |
|---|---|---|---|---|---|---|---|---|---|---|---|---|
| BGL | 0.089 | 0.126 | -0.303 | 0.052 | -0.028 | -0.662 | -0.711 | 1.000 | -0.901 | -0.677 | -0.473 | -0.766 |
| CCF | 0.397 | 0.386 | 0.648 | -0.245 | -0.214 | 0.235 | 0.463 | 0.033 | -0.026 | 0.126 | -0.239 | 0.318 |
| BCR | 0.567 | 0.409 | 0.408 | -0.193 | -0.158 | 0.175 | 0.283 |  | -0.748 | -0.169 | 0.138 | -0.257 |
| CPP | -0.496 | -0.024 | -0.052 | -0.158 | 0.480 | 0.709 | 0.369 | 0.034 | 0.898 | 0.597 | 0.283 | 0.247 |
| GSD | 0.066 | -0.071 | 0.488 | 0.129 | 0.293 | 0.276 | 0.720 | 0.733 | -0.079 | 0.480 | 0.422 | 0.553 |
| RBT | 0.091 | 0.246 | 0.251 | -0.277 | -0.013 | 0.126 | 0.043 | 0.098 | 0.316 | 0.111 | -0.100 | -0.073 |
| SMB | 0.943 | 0.963 | 0.541 | -0.375 | -0.509 | -0.306 | 0.080 | 0.845 | -0.422 | -0.265 | -0.149 | 0.327 |
| SPB | 0.785 | 0.776 | 0.477 | -0.606 | -0.391 | 0.097 | 0.081 | 0.694 | -0.013 | -0.440 | -0.326 | 0.211 |
| SXW | -0.696 | 0.145 | 0.308 | 0.291 | 0.696 | 0.336 | 0.719 | 0.699 | 0.813 | 0.853 | 0.710 | 0.775 |
| WAL | -0.543 | -0.206 | 0.070 | 0.385 | 0.884 | 0.663 | 0.532 | -0.496 | 0.796 | 0.718 | 0.468 | 0.390 |
| WCR | 0.059 | 0.602 | 0.978 | -0.082 | -0.162 | -0.021 | 0.344 | 0.784 | 0.045 | 0.294 | -0.247 | 0.395 |
| WHS | -0.370 | 0.071 | 0.352 | 0.285 | 0.441 | 0.479 | 0.698 | 0.910 | 0.075 | 0.734 | 0.585 | 0.623 |
| YPE | 0.636 | 0.191 | 0.001 | 0.188 | 0.018 | -0.411 | -0.813 | 1.000 | -0.985 | -0.697 | -0.039 | -0.926 |
| SMB < S | 1.000 | 0.809 | 0.323 | -0.724 | -0.750 | -0.111 | -0.216 | 1.000 | -0.950 | -0.778 | -0.512 | -0.167 |
| SMB S | 0.809 | 1.000 | 0.623 | -0.380 | -0.297 | -0.196 | 0.161 | 0.783 | -0.357 | 0.255 | -0.075 | 0.291 |
| SMB QP | 0.323 | 0.623 | 1.000 | -0.038 | -0.104 | -0.018 | 0.254 | 0.724 | -0.220 | 0.051 | -0.302 | 0.277 |
| WAL < S | -0.724 | -0.380 | -0.038 | 1.000 | 0.581 | -0.068 | 0.005 | -0.773 | 0.539 | 0.174 | 0.306 | 0.245 |
| WAL S | -0.750 | -0.297 | -0.104 | 0.581 | 1.000 | 0.516 | 0.482 | -0.547 | 0.785 | 0.698 | 0.677 | 0.465 |
| WAL Q | -0.111 | -0.196 | -0.018 | -0.068 | 0.516 | 1.000 | 0.426 | -0.683 | 0.634 | 0.418 | 0.259 | 0.129 |
| WAL PM | -0.216 | 0.161 | 0.254 | 0.005 | 0.482 | 0.426 | 1.000 | 0.513 | 0.283 | 0.812 | 0.634 | 0.784 |
| SXW < S | 1.000 | 0.783 | 0.724 | -0.773 | -0.547 | -0.683 | 0.513 | 1.000 | -0.181 | 0.765 | 0.949 | 0.765 |
| SXW S | -0.950 | -0.357 | -0.220 | 0.539 | 0.785 | 0.634 | 0.283 | -0.181 | 1.000 | 0.622 | 0.651 | 0.610 |
| SXW Q | -0.778 | 0.255 | 0.051 | 0.174 | 0.698 | 0.418 | 0.812 | 0.765 | 0.622 | 1.000 | 0.827 | 0.540 |
| SXW P | -0.512 | -0.075 | -0.302 | 0.306 | 0.677 | 0.259 | 0.634 | 0.949 | 0.651 | 0.827 | 1.000 | 0.729 |
| SXW M | -0.167 | 0.291 | 0.277 | 0.245 | 0.465 | 0.129 | 0.784 | 0.765 | 0.610 | 0.540 | 0.729 | 1.000 |

*Table A3: Correlation coefficients between annual relative weights for thirteen species sampled from Pueblo Reservoir, Colorado from 1990 to 2011 with different length classes of smallmouth bass, walleye, and hybrid striped bass. Color codes represent range of p-values. Magenta represents $p < 0.01$; green, $p < 0.05$; yellow, $p < 0.1$; red, $p < 0.2$.*

|  | SMB <S | SMB S | SMB QP | WAL <S | WAL S | WAL Q | WAL PM | SXW <S | SXW S | SXW Q | SXW P | SXW M |
|---|---|---|---|---|---|---|---|---|---|---|---|---|
| BGL | 0.417 | 0.394 | 0.233 | 0.447 | 0.474 | 0.026 | 0.024 |  | 0.018 | 0.033 | 0.099 | 0.008 |
| CCF | 0.145 | 0.120 | 0.016 | 0.210 | 0.241 | 0.220 | 0.065 | 0.484 | 0.478 | 0.348 | 0.205 | 0.134 |
| BCR | 0.217 | 0.296 | 0.248 | 0.378 | 0.375 | 0.389 | 0.408 |  | 0.231 | 0.300 | 0.412 | 0.338 |
| CPP | 0.072 | 0.472 | 0.440 | 0.303 | 0.049 | 0.003 | 0.119 | 0.483 | 0.003 | 0.020 | 0.163 | 0.197 |
| GSD | 0.428 | 0.418 | 0.064 | 0.337 | 0.166 | 0.181 | 0.004 | 0.134 | 0.433 | 0.057 | 0.066 | 0.020 |
| RBT | 0.415 | 0.262 | 0.257 | 0.219 | 0.485 | 0.364 | 0.450 | 0.451 | 0.245 | 0.380 | 0.385 | 0.416 |
| SMB | 0.000 | 0.000 | 0.053 | 0.103 | 0.038 | 0.167 | 0.407 | 0.075 | 0.173 | 0.220 | 0.314 | 0.138 |
| SPB | 0.006 | 0.002 | 0.069 | 0.024 | 0.131 | 0.395 | 0.418 | 0.255 | 0.489 | 0.118 | 0.164 | 0.267 |
| SXW | 0.019 | 0.335 | 0.178 | 0.167 | 0.004 | 0.131 | 0.004 | 0.155 | 0.013 | <0.001 | 0.002 | 0.001 |
| WAL | 0.065 | 0.272 | 0.419 | 0.097 | < 0.001 | 0.007 | 0.038 | 0.252 | 0.016 | 0.004 | 0.046 | 0.384 |
| WCR | 0.450 | 0.057 | 0.000 | 0.417 | 0.351 | 0.479 | 0.202 | 0.213 | 0.471 | 0.221 | 0.261 | 0.146 |
| WHS | 0.164 | 0.418 | 0.144 | 0.173 | 0.066 | 0.058 | 0.009 | 0.045 | 0.436 | 0.010 | 0.018 | 0.012 |
| YPE | 0.124 | 0.379 | 0.499 | 0.381 | 0.484 | 0.209 | 0.047 |  | 0.015 | 0.062 | 0.471 | 0.004 |
| SMB < S |  | 0.004 | 0.198 | 0.014 | 0.010 | 0.388 | 0.304 | 0.500 | 0.007 | 0.007 | 0.079 | 0.334 |
| SMB S | 0.004 |  | 0.027 | 0.125 | 0.188 | 0.282 | 0.340 | 0.109 | 0.216 | 0.339 | 0.413 | 0.193 |
| SMB QP | 0.198 | 0.027 |  | 0.456 | 0.380 | 0.479 | 0.255 | 0.242 | 0.318 | 0.448 | 0.183 | 0.205 |
| WAL < S | 0.014 | 0.125 | 0.456 |  | 0.451 | 0.368 | 0.226 | 0.138 | 0.106 | 0.304 | 0.155 | 0.210 |
| WAL S | 0.010 | 0.188 | 0.380 | 0.451 |  | 0.066 | 0.067 | 0.170 | 0.018 | 0.009 | 0.006 | 0.055 |
| WAL Q | 0.388 | 0.282 | 0.479 | 0.368 | 0.066 |  | 0.096 | 0.159 | 0.063 | 0.088 | 0.186 | 0.330 |
| WAL PM | 0.304 | 0.340 | 0.255 | 0.226 | 0.067 | 0.096 |  | 0.243 | 0.261 | 0.001 | 0.013 | 0.001 |
| SXW < S | 0.500 | 0.109 | 0.242 | 0.138 | 0.170 | 0.159 | 0.243 |  | 0.385 | 0.118 | 0.026 | 0.118 |
| SXW S | 0.007 | 0.216 | 0.318 | 0.106 | 0.018 | 0.063 | 0.261 | 0.385 |  | 0.068 | 0.055 | 0.073 |
| SXW Q | 0.007 | 0.339 | 0.448 | 0.304 | 0.009 | 0.088 | 0.001 | 0.118 | 0.068 |  | <0.001 | 0.035 |
| SXW P | 0.079 | 0.413 | 0.183 | 0.155 | 0.006 | 0.186 | 0.013 | 0.026 | 0.055 | <0.001 |  | 0.003 |
| SXW M | 0.334 | 0.193 | 0.205 | 0.210 | 0.055 | 0.330 | 0.001 | 0.118 | 0.073 | 0.035 | 0.003 |  |

*Table A4: p-values for correlation coefficients shown in Table A3.*

**Acknowledgments**

Funding was provided in part by BTG Research (www.BTGResearch.org) and the United States Air Force Academy. The authors are grateful to Colorado Division of Parks and Wildlife for providing survey data. Valuable feedback on the project and a manuscript draft was provided by Michelle McGree (CDPW), David Willis (SDSU), and Michael Blackwell (South Dakota Department of Game, Fish, and Parks). All inaccuracies and errors are the responsibility of the authors.